%% file: main.tex
\documentclass[acmlarge,screen,nonacm]{acmart}

\setcopyright{none}
\renewcommand\footnotetextcopyrightpermission[1]{} 
\AtBeginDocument{%
  \providecommand\BibTeX{{%
    \normalfont B\kern-0.5em{\scshape i\kern-0.25em b}\kern-0.8em\TeX}}}

\usepackage{hyperref}
\usepackage{caption}
\usepackage{subcaption}
\usepackage[noend]{algpseudocode}
\usepackage{algorithm}
\usepackage{mathtools}
\usepackage{amsmath}
\usepackage{graphicx}
\usepackage{textcomp}
\usepackage{multirow}
\usepackage{tikz}
\usepackage{enumitem}
\usepackage{booktabs}
\usepackage{soul}

\newcommand{\ie}{{\em i.e.}}
\newcommand{\eg}{{\em e.g.}}
\newcommand{\dataDependent}{{sparsity-dependent }}

\newcommand{\tool}{{Sparseloop}}
\DeclareRobustCommand\circled[1]{\tikz[baseline=(char.base)]{
            \node[shape=circle,draw,inner sep=0.5pt](char){#1};}}

\begin{document}

\title{\tool: An Analytical Approach To Sparse Tensor Accelerator Modeling}


\author{Yannan Nellie Wu}
\email{nelliewu@mit.edu}
\affiliation{
  \institution{MIT}
  \streetaddress{}
  \city{Cambridge}
  \state{MA}
  \country{USA}
}

\author{Po-An Tsai}
\email{poant@nvidia.com}
\affiliation{
  \institution{NVIDIA}
  \streetaddress{}
  \city{Westford}
  \state{MA}
  \country{USA}
}

\author{Angshuman Parashar}
\email{aparashar@nvidia.com}
\affiliation{
  \institution{NVIDIA}
  \streetaddress{}
  \city{Westford}
  \state{MA}
  \country{USA}
}

\author{Vivienne Sze}
\email{sze@mit.edu}
\affiliation{
  \institution{MIT}
  \streetaddress{}
  \city{Cambridge}
  \state{MA}
  \country{USA}
}

\author{Joel S. Emer}
\email{emer@csail.mit.edu}
\affiliation{
  \institution{MIT/NVIDIA}
  \streetaddress{}
  \city{Cambridge}
  \state{MA}
  \country{USA}
}

\begin{abstract}
In recent years, many accelerators have been proposed to efficiently process sparse tensor algebra applications (\eg, sparse neural networks). However, these proposals are single points in a large and diverse design space. The lack of systematic description and modeling support for these sparse tensor accelerators impedes hardware designers from efficient and effective design space exploration.

This paper first presents a unified taxonomy to systematically describe the diverse sparse tensor accelerator design space. Based on the proposed taxonomy, it then introduces \tool, the first fast, accurate, and flexible analytical modeling framework to enable early-stage evaluation and exploration of sparse tensor accelerators. \tool~comprehends a large set of architecture specifications, including various dataflows and sparse acceleration features (\eg, elimination of zero-based compute). Using these specifications, \tool~evaluates a design's processing speed and energy efficiency while accounting for data movement and compute incurred by the employed dataflow, including the savings and overhead introduced by the sparse acceleration features using stochastic  density models.

Across representative accelerator designs and workloads, \tool~achieves over 2000$\times$ faster modeling speed than cycle-level simulations, maintains relative performance trends, and achieves 0.1\% to 8\% average error. The paper also presents example use cases of \tool~in different accelerator design flows to reveal important design insights.

\end{abstract}

\pagestyle{plain}

\maketitle

\input{01_introduction}

\input{02_motivation}

\input{03_classification}

\input{04_overview}
\input{05_0_inputs}

\input{05_1_dataflow_modeling}

\input{05_2_sparse_modeling}

\input{05_3_microarch_modeling}

\input{06_1_speed}

\input{06_2_validation}
\input{06_4_stc_explore}

\input{06_3_case_study}

\input{07_related_work}
\input{08_conclusion}
\begin{acks}
We thank Haoquan Zhang for discussions on statistical analysis of tensor density. We would also like to thank the anonymous reviewers for their constructive feedback. 

Part of the research was done during Yannan Nellie Wu’s internship at NVIDIA Research. This research was funded in part by the U.S. Government. The views and conclusions contained in this document are those of the authors and should not be interpreted as representing the official policies, either expressed or implied, of the U.S. Government. This research was also funded in part by DARPA contract HR0011-18-3-0007, NSF PPoSS 2029016 and Ericsson. 
\end{acks}

\input{10_artifact_appendix.tex}

\newpage
\bibliographystyle{ACM-Reference-Format}
\bibliography{reference}

\end{document}

%% file: 01_introduction.tex
\section{Introduction}
Sparse tensor algebra is widely used in many important applications, such as scientific simulations~\cite{power-simulation}, computer graphics~\cite{graphics}, graph algorithms~\cite{sparse-graph, social-networks-graph}, and deep neural networks (DNNs)~\cite{tensorflow, pytorch}. Depending on the sparsity characteristics of the tensors (\eg, sparsity, distribution of zero locations), sparse tensor algebra can introduce a significant number of \emph{ineffectual computations}, whose results can be easily derived by applying the simple algebraic equalities of $X\times 0 = 0$ and $X + 0 = X$, without reading all the operands or doing the computations~\cite{ineffectual-work, extensor}.

As a result, many performant and energy-efficient sparse tensor algebra accelerators have been proposed to exploit ineffectual computations to reduce data movement and compute~\cite{eyeriss-v1, eyeriss-v2, scnn, extensor, cnvlutin, matraptor, OuterSPACE, SIGMA, GAMMA-accelerator, SparTen, ampere-white-paper, samsung-npu, GoSPA, dual-side-tensor-core}. Based on the properties of its target applications (\eg, convolution or matrix multiplication), each accelerator design proposes its unique hardware support. Various accelerators can: propose different architecture topology (\eg, number of storage levels); employ different \emph{dataflows}~\cite{eyeriss-v1} (\ie, the rules for scheduling data movement and compute in space and time); use different \emph{encoding} to compress sparse tensors (\eg, bitmask encoding); and design different hardware
to eliminate operations associated with ineffectual computations (\eg, intersection units).
The joint design space for all these hardware mechanisms is therefore large and diverse.

To characterize either a single specific design or many designs as part of design space exploration, hardware designers can benefit from a modeling framework that is:
\begin{itemize}
\item \emph{\textbf{Flexible:}} is capable of modeling a diverse range of potential designs with hardware support for different dataflows, compression encodings, etc.
\item \emph{\textbf{Fast:}} produces simulation results quickly. This is particularly important because properly characterizing a specific design requires finding the best schedule, i.e., \emph{mapping}, for a given workload, which generally requires a search of a large mapspace~\cite{gamma, marvel, timeloop-ispass}.
\item \emph{\textbf{Accurate:}} produces simulation results correctly in both mapspace and design space exploration.
\end{itemize}

However, to the authors' knowledge, none of the existing modeling approaches for tensor accelerators provide the desired capabilities (Table~\ref{tab:existing_work}). 
On the one hand, \emph{cycle-level design-specific models}~\cite{eyeriss-v1, eyeriss-v2, scnn, extensor, cambriconX, cnvlutin, sparch, Asilomar, matraptor, OuterSPACE, SIGMA, STONNE} capture the detailed implementations for their target designs (\eg, memory control signals) and thus are very accurate. 
However, such models impede users from mapspace exploration due to their slow simulation speed and design space exploration due to their limited parameterization support.
On the other hand, analytical models perform mathematical computations to analyze the important high-level characteristics of a class of accelerators and are fast. However, the existing \emph{general analytical models} are only flexible for dense accelerator designs~\cite{timeloop-ispass, cosa, accelergy-ispass, scale-sim, maestro, nnest-islped, DNN_predictor, DnnWeaver, AutoDNNchip, T2STensor}, \ie, they do not reflect the impact of sparsity-aware acceleration techniques, resulting in inaccurate modeling.

\input{tables/exisiting_work}
To address the limitations of existing work, we present \tool\footnote{\tool~is open-source and publicly available at \href{http://sparseloop.mit.edu/}{Sparseloop website}~\cite{sparseloop-website}.}, 
the \emph{first analytical modeling framework} for fast, accurate, and flexible evaluations of sparse (and dense) tensor accelerators, enabling early-stage exploration of the large and diverse sparse tensor accelerator design space. Table~\ref{tab:existing_work} compares \tool~to existing simulation frameworks.

\noindent\textit{\textbf{This work makes the following key contributions:}}

\noindent\textbf{(1)} To systematically describe the large and diverse sparse tensor accelerator design space, we propose a taxonomy to classify the various sparsity-aware acceleration techniques into three \emph{sparse acceleration features (SAFs)}: representation format, gating, and skipping.

\noindent\textbf{(2)} Based on the SAF classification, we propose \tool, an analytical modeling framework for tensor accelerators.
\begin{itemize}
    \item To both faithfully reflect workload data's impact on accelerator performance and ensure simulation speed, \tool~performs analysis based on statistical characterizations of nonzero value locations in the tensors. 
    \item To keep the modeling complexity tractable and allow support for emerging workloads/designs, \tool~splits its modeling process into discrete steps, each of which focuses on evaluating a distinct design aspect (\eg, dataflow, sparse acceleration features). This decoupling allows modeling of both dense and sparse designs in one infrastructure.
\end{itemize}

\noindent\textbf{(3)} With representative accelerator designs and workloads, we show that \tool~is fast, accurate, and flexible:
\begin{itemize}
    \item \tool~runs more than 2000$\times$ faster than a cycle-level simulator.
    \item \tool~maintains relative performance trends and achieves 0.1\% to 8\% average error across designs.
    \item \tool~allows comparison of designs with different dataflows and sparse acceleration features, running workloads with various sparsity characteristics. 
    \item \tool~can reveal design insights during  accelerator design flows. Our case studies demonstrate \tool's flexibility to quickly compare and explore diverse designs with different architectures, dataflows, and SAFs running various workloads.
 
\end{itemize}

%% file: tables/exisiting_work.tex
\begin{table}[tb]
\centering
\resizebox{0.6\columnwidth}{!}{
\begin{tabular}{c|c|c|c|c}
\begin{tabular}[c]{@{}c@{}}\\  \end{tabular} &
  \textbf{Accuracy} &
  \textbf{Speed} &
  \textbf{Flexibility} &
  \begin{tabular}[c]{@{}c@{}}\textbf{Support} \\ \textbf{Sparsity?}\end{tabular} \\ \hline \hline \\ [-1em]
\begin{tabular}[c]{@{}c@{}}Cycle-Level \\ Design-Specific\end{tabular} & \textbf{Very High} & \textbf{Slow} & \textbf{Low}  & \textbf{Yes} \\ \hline \rule{0pt}{10pt}
\begin{tabular}[c]{@{}c@{}}General Analytical\end{tabular}       & \textbf{High} & \textbf{Fast} & \textbf{High} & \textbf{No}  \\ \hline \rule{0pt}{10pt}
\textbf{\begin{tabular}[c]{@{}c@{}} Our Work\end{tabular}}    & \textbf{High} & \textbf{Fast} & \textbf{High} & \textbf{Yes} \\
\hline
\end{tabular}
}

\caption{Comparison of existing tensor accelerator simulation frameworks with our proposed \tool~framework.}
\vspace{-5pt}

\label{tab:existing_work}
\end{table}




%% file: 02_motivation.tex
\section{Background and Motivation}
\label{sec:movtivation}


In this section, we illustrate the complexity of describing and evaluating the sparse tensor accelerator design space.

\subsection{Large and Unstructured Design Space}
 Sparse tensor accelerators often employ different dataflows to exploit data reuse across multiple storage levels 
 and feature various sparsity-aware acceleration techniques to eliminate data storage for zeros and \emph{ineffectual operations} (IneffOps), \ie, arithmetic operations and storage accesses associated with ineffectual computations. The vast number of potential design choices lead to a large and diverse design space.
 
 Nonetheless, there is little structure in the design space for sparse tensor accelerators,
 as each prior design uses different terminology to describe a point in the design space. 
 We present the design decisions made by representative designs to show the lack of uniformity in their architecture proposals.
 
 For example, Eyeriss~\cite{eyeriss-v1} uses a \emph{row-stationary} dataflow, a \emph{RLC} encoding for data stored in DRAM, and storage and compute units that stay idle for IneffOps. With the same dataflow, Eyeriss V2~\cite{eyeriss-v2} employs a \emph{compressed sparse column} encoding for both on-chip and DRAM data, and avoids spending cycles for IneffOps by performing intersections near the compute units. 
 SCNN~\cite{scnn} also uses a similar intersection-based acceleration, but features a \emph{PlanarTiled-InputStationary-CartesianProduct} dataflow and \emph{compressed-sparse-block} encoding. ExTensor~\cite{extensor} proposes a \emph{hybrid} dataflow and a \emph{hierarchical} encoding. It introduces the \emph{hierarchical-elimination} acceleration technique, which aggressively eliminates IneffOps at multiple storage levels long before data reaches compute. Dual-side sparse tensor core (DSTC)~\cite{dual-side-tensor-core} uses an \emph{output-stationary} dataflow and \emph{two-level BitMap} encoding. It designs an \emph{operand-collector} hardware unit tailored to its dataflow to provide enough bandwidth after elimination of IneffOps.
 
 
 Since different accelerators propose different sets of implementation choices, often described in design-specific naming conventions, it is challenging for  designers to have a systematic understanding of the proposed dataflow and acceleration techniques in the design space, let alone a modeling framework to compare these designs systematically.

\subsection{Sparsity Impacts Design Behavior}

 
 Evaluating the complex design space of sparse tensor accelerators is further complicated by the impact of the tensor \emph{sparsity characteristics}, which include the density (\ie, percentage of nonzero values in each tensor, 1$-$sparsity) and the locations of nonzero values in each tensor.
 
 To demonstrate this entanglement, we compare two 
 designs supporting different data representations. For simplicity, both accelerators employ the same dataflow:
 
\begin{figure}[tb]
	\centering
    \includegraphics[width=0.8\linewidth]{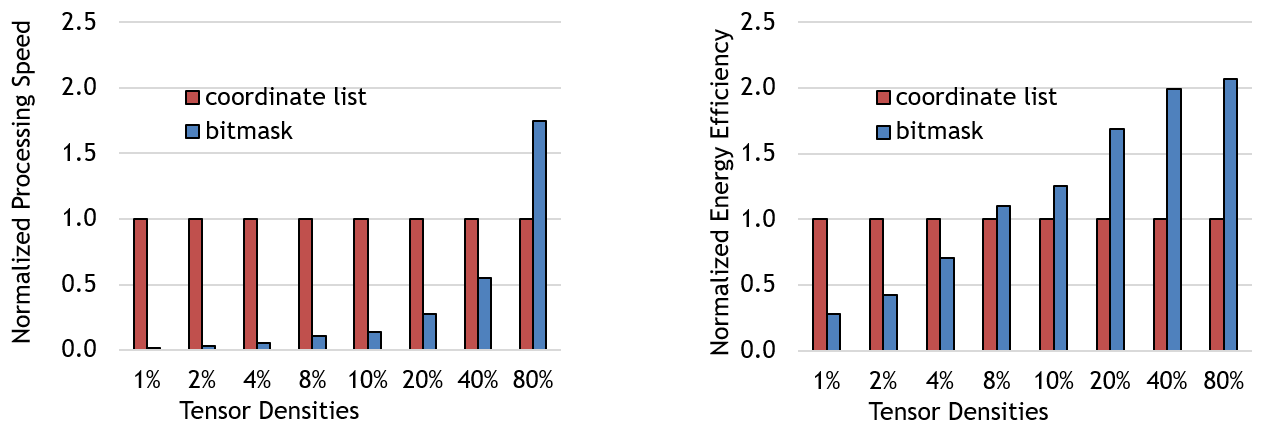}
    \caption{Processing speed and energy efficiency of architectures with different data representation support running sparse matrix multiplication workloads. Design behavior is dependent on data representations and tensor densities.}
    \label{fig:mov.diff_arch}
\end{figure}

\noindent\textbf{(1) Bitmask (Eyeriss-like)}: The first design supports bitmask encoding to represent sparse operand tensors. Bitmask uses a single bit to encode whether each value is nonzero or not. In each cycle, the design uses each bit to decide whether its storage and compute units should stay idle to save energy, but it does not improve processing speed.\footnote{Of course, there exist other designs that use bitmasks to save both energy and time ~\cite{SparTen, cambriconX}}

\noindent\textbf{(2) Coordinate list (SCNN-like)}: The second design employs a coordinate list encoding~\cite{compression-formats, dnn_syn_lec}, which indicates the location of each nonzero value via a list of its coordinates (\ie, the indices in each dimension). Since the coordinate information directly points to the next effectual computation, the design only spends cycles on effectual operations, thus saving both energy and time.

In Fig.~\ref{fig:mov.diff_arch}, we compare the processing speed and energy efficiency of the two designs running sparse matrix multiplication workloads of different densities. 
As shown in Fig.~\ref{fig:mov.diff_arch}, the best design choice is a function of the input density. More specifically, since the bitmask-based design does not improve processing speed, with low-density tensors, \emph{bitmask} always runs slower than \emph{coordinate list}. However, since \emph{coordinate list} needs to encode the exact coordinates with multiple bits, it incurs more significant encoding overhead per nonzero value. As the tensors become denser, \emph{coordinate list} leads to lower energy efficiency and/or processing speed. 
This trend has also been observed in Sigma~\cite{SIGMA}.

Even just varying the input tensor density, we already see non-trivial interactions between the benefits introduced by eliminated IneffOps and compressed sparse tensors, and the overhead introduced by extra encoding information. 
A more involved case study in Sec.~\ref{sec:stc-study} will further showcase the complex interactions between dataflows, sparsity-aware acceleration techniques, and workload sparsity characteristics, illustrating the importance of co-designing various design aspects. Thus, for hardware designers to efficiently explore the trade-offs of various design decisions, there is a strong need to have a \emph{fast modeling framework that, in addition to evaluating various dataflows, recognizes the impact of the different acceleration techniques and tensor sparsity characteristics on processing speed and energy efficiency}.

\noindent\textbf{Our proposal:}
To address these two issues, we first introduce a new classification of the various sparsity-aware acceleration techniques (Sec.~\ref{sec:classification}), which
unifies how to qualitatively describe these techniques in a systematic manner.
Leveraging this 
taxonomy, we then propose an analytical modeling framework, \tool, which quantitatively evaluates the diverse sparse tensor accelerator designs 
(Sec.~\ref{sec:high_level} to Sec.~\ref{sec:microarch}).
We show that \tool~is fast, accurate, and flexible in Sec.~\ref{sec:exp_speed},~\ref{sec:exp_val}, and~\ref{sec:case-study}.

%% file: 03_classification.tex
\section{Design Space Classification} \label{sec:classification}

The first step toward a systematic modeling framework is to have a unified taxonomy
to describe various sparse tensor accelerators. We propose a new classification framework that simplifies how to describe a specific design in the complex design space. We then demonstrate how prior designs can be described in a straightforward manner.

\subsection{High-Level Sparse Acceleration Features}
\label{sec:SAFs}
To systematically describe sparse tensor accelerators in the design space, we classify common sparsity-aware acceleration techniques into three orthogonal high-level categories: 
\begin{itemize}
    \item Representation format
    \item Gating IneffOps
    \item Skipping IneffOps
\end{itemize}
We call each category a \emph{sparse acceleration feature} (SAF). 




\subsubsection{Representation Format}
Representation format refers to the choice of encoding the locations of nonzero values in the tensor. To describe a representation format, we adopt a hierarchical expression that combines multiple per-dimension formats, similar to~\cite{taco, compression-formats, dnn_syn_lec}. 

\begin{figure}[tb]
     \centering
     \includegraphics[width=0.7\linewidth]{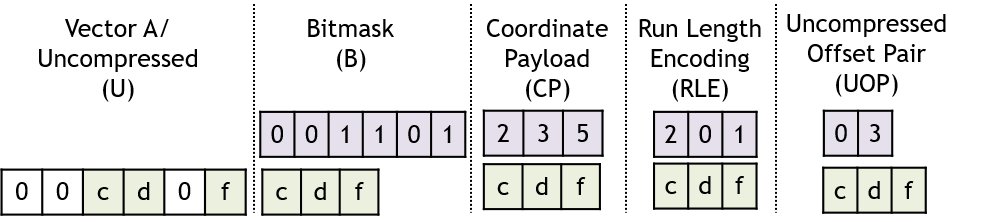}
     \caption{Example representation formats of a vector A. Purple vectors refer to metadata used to identify the original locations of the nonzero values.}
    \label{fig:format}
\end{figure}

As shown in Fig.~\ref{fig:format}, we introduce several commonly used per-dimension formats with an example 1D tensor, \ie, a vector. The most basic format is \emph{Uncompressed (U)}, which represents the tensor with its exact values, thus directly showing the locations of nonzero values. \textit{U} is identical to the original vector. However, to save storage space, and thus implicitly save energy (and time) associated with zero value accesses, sparse tensor accelerators tend to employ \emph{compressed formats}, which represent a tensor with only nonzero values and some additional information about their original locations or \emph{coordinate}~\cite{taco, compression-formats, dnn_syn_lec}. We call this information \emph{metadata}. We introduce four per-dimension compressed formats\footnote{Of course, many more per-dimension formats exist
and can be incorporated modularly into~\tool.}. 

\begin{itemize}
    \item \emph{Coordinate Payload (CP)}:  
    the coordinates of each nonzero value are encoded with multiple bits.
    The \textit{payloads} are either the nonzero value or a pointer to another dimension. CP explicitly lists the coordinates and the corresponding payloads.
    
    \item \emph{Bitmask (B)}: a single bit is used to encode whether each coordinate is nonzero or not.
    
    \item \emph{Run Length Encoding (RLE)}: multiple bits are used to encode the run length, which represents the number of zeros between nonzeros (\eg, an $r$-bit run length can encode up to a $2^r-1$ run of zeros).
    
    \item \emph{Uncompressed Offset Pairs (UOP)}: multiple bits are used to encode the start (inclusive) and end (noninclusive) positions of nonzero values.
\end{itemize}

\begin{figure*}
     \begin{subfigure}[b]{0.14\linewidth}
         \centering
         \includegraphics[width=\linewidth]{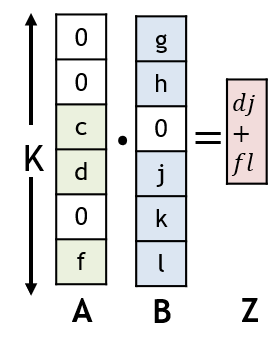}
         \caption{}
         \label{fig:workload}
     \end{subfigure}
     \hfill
     \begin{subfigure}[b]{0.82\linewidth}
         \centering
         \includegraphics[width=\linewidth]{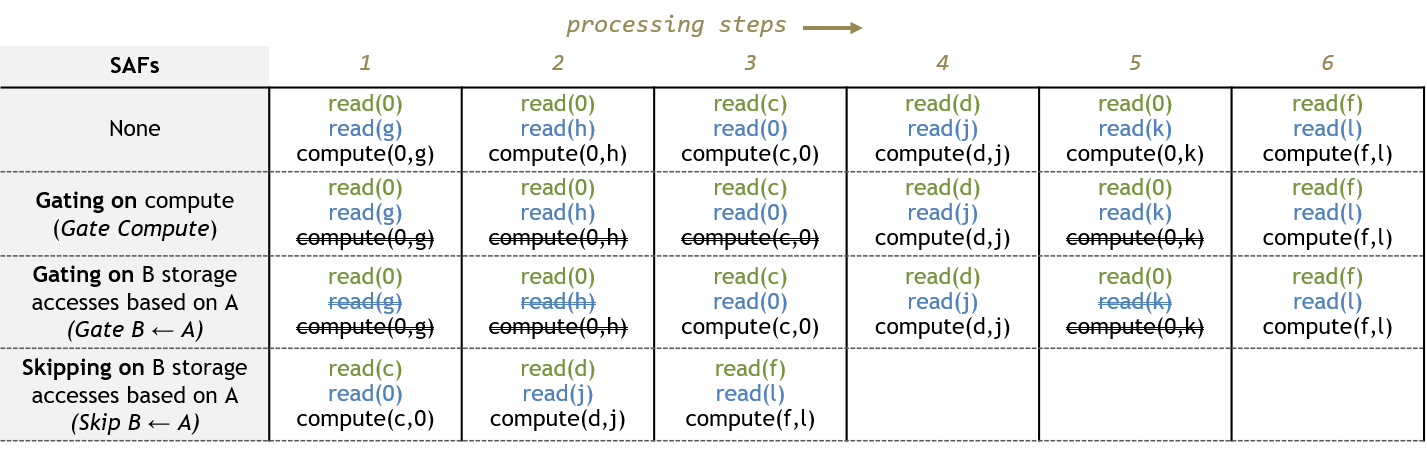}
         \caption{}
         \label{fig:sparse_features.gating_skipping}
     \end{subfigure}
     \vspace{10pt}
     \caption{(a) Sparse dot product workload. (b) Example ways of processing the example workload. 
     1st row: baseline processing without SAFs; 2nd row: Gating applied to compute; 3rd row: Gating applied to B reads based on A's values; 4th row: Skipping applied to B reads based on A's values. }
      \label{fig:workload.gating_skipping}
\end{figure*}

\input{tables/example_formats}
As shown in Table~\ref{tab:exampe_formats}, full tensor representation formats can be described by combing the per-dimension formats in a hierarchical fashion. For example, CSR (compressed sparse row)~\cite{CSR} can be described by \emph{UOP-CP}: Top level UOP encodes the start and end locations of the nonzeros in each row; bottom level CP encodes the exact column coordinates and its associated nonzeros. A format can also split and/or flattened tensor dimensions (\eg, 2D COO flattens multiple dimensions into one dimension represented by CP with tuples as coordinates). We use a superscript to indicate the number of flattened dimensions.

\subsubsection{Gating}
Gating exploits the existence of IneffOps by letting the storage and compute units stay idle during the corresponding cycles. As a result, it saves energy but does not change processing speed. Gating can be applied to both compute and storage units in the architecture.

We use the dot-product workload in Fig.~\ref{fig:workload} to illustrate the impact of gating. Each row of Fig.~\ref{fig:sparse_features.gating_skipping} corresponds to a specific SAF implementation, each column is a processing step, and each cell lists the operations happening at the step. The first row presents the baseline processing without any SAFs applied, so it performs all IneffOps and takes six steps to complete.

The second row in Fig.~\ref{fig:sparse_features.gating_skipping} shows the result of applying gating to compute units, $Gate\>Compute$. The compute unit checks whether operands are zeros and stays power-gated if at least one operand is zero.

When gating is applied to storage units, it can be based on one of two approaches:

\noindent\textbf{\emph{(1) Leader-follower intersection}} checks one operand, and if this operand is zero, it avoids accessing the other operand. We call the checked operand the \emph{leader} and the operand with gated access the \emph{follower}. In our classification, gating based on leader-follower intersection is represented by an arrow that points from the leader to the follower, \ie, $Gate\> Follower\leftarrow Leader$. The third row in Fig.~\ref{fig:sparse_features.gating_skipping} shows $Gate \> B \leftarrow A$. Note that this approach may not eliminate all IneffOps (\eg, step three in the example), and the savings introduced depend on the leader operand's sparsity characteristics.

\noindent\textbf{\emph{(2) Double-sided intersection}} checks both operands (usually just via their associated metadata), and if either of them is zero, it does not access either operands' data. Double-sided intersection is represented with a double-sided arrow that points to both operands, \ie, $Operand0\leftrightarrow Operand1$. Double-sided intersection eliminates all IneffOps but may require more complex hardware.

 In addition to reducing storage accesses, gating applied to a storage unit also leads to \emph{implicit} gating of the compute unit connected to it (\eg, step one in the third row of Fig.~\ref{fig:sparse_features.gating_skipping}), as the compute unit can now use the check for the storage unit to power-gate itself.

  
\subsubsection{Skipping} 
Skipping refers to exploiting IneffOps by not spending the corresponding cycles. Since skipping directly skips to the next effectual computation, it saves both energy and time. Similar to gating, skipping can be applied to both the compute and storage units. 

When skipping is applied to compute units, the compute units directly look for the next pair of operands until it finds effective computations to perform. When skipping is applied to storage units, it can also be based on leader-follower intersection or double-sided intersection. However, instead of letting the storage stay idle, with skipping applied, cycles are only spent on effectual accesses. The last row in Fig.~\ref{fig:sparse_features.gating_skipping} shows an example implementation of skipping B reads based on A's values ($Skip \> B \leftarrow A$).  Similar to gating, a leader-follower implementation of skipping can still introduce some IneffOps, and skipping at storage can lead to implicit skipping at the compute units. Since skipping needs to quickly locate the next effectual operation to skip to, it usually requires more complex hardware than gating does (\eg, ExTensor's intersection unit implements smart look-ahead optimizations to locate effectual operations in time~\cite{extensor}). Inefficient implementations can lead to more overhead than savings in time and energy.

\input{tables/arch_summary}
\subsection{Dataflow is Orthogonal to Sparsity-Aware Acceleration}

In addition to the SAFs, dataflow choice is another important decision made by various accelerators~\cite{eyeriss-isca}. A taxonomy of dataflows for various tensor algebra workloads has already been well studied in existing work (\eg, for DNNs~\cite{eyeriss-isca, dnn_syn_lec}, and matrix multiplications~\cite{GAMMA-accelerator, matraptor, OuterSPACE}).

We make the observation that the dataflow choice is \emph{orthogonal to} the chosen SAFs. Dataflows define the scheduling of data movement and compute in time and space, and SAFs define the actual amount of data that is moved or number of computes performed.
As a result, the space of sparse tensor accelerators is the cross product of dataflow choices and SAF choices (further information on how this impacts modeling is in Section~\ref{sec:high_level}). Of course, a particular dataflow might mesh well with a specific SAF implementation, leading to an efficient design, while another may not.

\subsection{Describing Sparse Tensor Accelerators}

\begin{figure}[tb]
	\centering
	\includegraphics[width=0.6\linewidth]{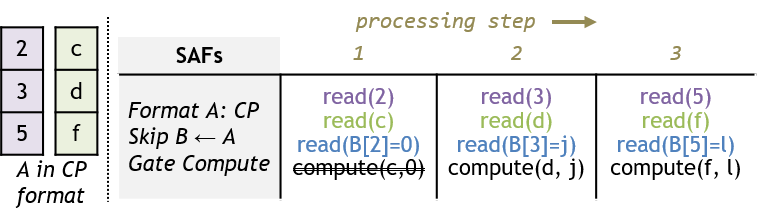}
	\caption{Example of combining compressed format, skipping, and gating SAFs in one design.}
	\vspace{-4pt}
	\label{fig:compounded_features}
	
\end{figure}
General accelerator designs often implement multiple SAFs that work well with each other to efficiently improve hardware performance. Fig.~\ref{fig:compounded_features} illustrates the idea with a simple example, for the same workload in Fig.~\ref{fig:workload},  Fig.~\ref{fig:compounded_features} employs a CP representation format for vector A, $Skip\>B \leftarrow A$ and $Gate\>Compute$. By representing A with CP, $Skip\> B \leftarrow A$ is implemented by directly reading the appropriate B values based on A's metadata. Furthermore, by applying $Gate\>Compute$, Fig.~\ref{fig:compounded_features} eliminates the compute unit's IneffOps for cases with nonzero A and zero B. 

Realistic sparse tensor accelerators often feature multiple storage levels to exploit data reuse opportunities and a set of spatial compute units for parallel computation. Thus, to systematically describe each design, we need to define the SAFs implemented at each level in the architecture. Based on our proposed classification, Table~\ref{tab:arch_summary} describes the acceleration techniques of the representative designs introduced in Sec.~\ref{sec:movtivation}.  

For example, SCNN~\cite{scnn}, a sparse DNN accelerator, uses a three-level, B-UOP-RLE representation format\footnote{Some per-dimension formats are applied to split or flattened dimensions.}  to compress input activation (IA) and weights (W). In the innermost storage level, \ie, the level closest to the compute units, SCNN performs $Skip W \> \leftarrow IA$ and $Skip\> OA \leftarrow IA\>\&W$, where OA refers to output activation.  Gating is applied to compute units, \ie, $Gate\>Compute$, to eliminate leftover IneffOps, similar to Fig.~\ref{fig:compounded_features}'s strategy. ExTensor~\cite{extensor} is an accelerator for general sparse tensor algebra. We use matrix multiplication as an example workload, which involves operand tensors A, B and result tensor Z. ExTensor partitions and compresses tensors with a six-level format and performs $Skip\>A\leftrightarrow B$ and $Skip\>Z\leftarrow A\&B$ at all storage levels.

Thus we hope it is clear how this taxonomy allows the design-specific terminologies in existing proposals to be translated into systematic descriptions. More importantly, it also allows future sparse accelerators to be described accurately and compared \emph{qualitatively} in the same way.

%% file: tables/example_formats.tex
\begin{table}[tb]
\centering

\begin{tabular}{c|c}
\begin{tabular}[c]{@{}c@{}}Example Classic \\ Representation Formats\end{tabular} & \multicolumn{1}{l}{\begin{tabular}[c]{@{}l@{}}Hierarchical \\ Description\end{tabular}} \\ \hline \hline \\[-1em]
Compressed Sparse Row (CSR)~\cite{CSR} & UOP-CP                                                                                                        \\
2D Coordinate List (COO)~\cite{COO}        & CP$^{2}$                                                                                                        \\
Compressed Sparse Block (CSB)~\cite{CSB}          & UOP-CP-CP     \\
3D Compressed Sparse Fibers (CSF)~\cite{CSF}       & CP-CP-CP
\end{tabular}

\caption{Example representation formats and their hierarchical description based on per-dimension formats.} 

\label{tab:exampe_formats}
\end{table}

%% file: tables/arch_summary.tex
\begin{table*}[ht]
\centering
\resizebox{\textwidth}{!}{
\begin{tabular}{c|c|l|l}

\textbf{Design}     & \textbf{Workload} & \multicolumn{1}{c|}{\textbf{Format$^{4}$}} & \multicolumn{1}{c}{\textbf{Gating/Skipping}}                          \\ \hline\hline \\[-1em]
    \textbf{Eyeriss~\cite{eyeriss-v1}}    & DNN                  & \begin{tabular}[c]{@{}l@{}}offchip:  I/O: B-RLE W:U \\ onchip:  I: UB O/W:U \end{tabular}          & Innermost Storage : $Gate\> W \leftarrow I$, $Gate\>  O \leftarrow I$   \\ \hline
\rule{0pt}{8pt}
\textbf{Eyeriss V2~\cite{eyeriss-v2}} & DNN                  & I/W: B-UOP-CP  O:U                  & Innermost Storage : $Skip\>  W \leftarrow I$, $Skip\>  O \leftarrow I\>\& W$; $Gate\>  Compute$ \\ \hline
\rule{0pt}{8pt}
\rule{0pt}{8pt}
\textbf{SCNN~\cite{scnn}}             & DNN                  & I/W: B-UOP-RLE  O: U          & Innermost Storage : $Skip\>  W \leftarrow I$, $Skip\>  O \leftarrow I \>\& W$; $Gate\>  Compute$ \\ \hline
\rule{0pt}{8pt}
\textbf{ExTensor~\cite{extensor}}     & MM & A/B: UOP-CP$\times5$   Z: U                   & All Storage : $Skip\>  A \leftrightarrow B$,  $Skip\>  Z \leftarrow A \>\& B$                     \\ \hline
\rule{0pt}{8pt}
\begin{tabular}[c]{@{}c@{}}\textbf{DSTC}~\cite{dual-side-tensor-core}\end{tabular}& MM & A/B: B-B   Z: U   & \begin{tabular}[l]{@{}l@{}} $2^{nd}$-to-innermost \& Innermost Storage :  $Skip\>  A \leftrightarrow B$,  $Skip\>  Z \leftarrow A \>\& B$    \end{tabular}                 \\
\bottomrule
\end{tabular}
}
\caption{Summary of representative sparse tensor accelerators described with the proposed SAFs based on tensors from example target workloads. For DNN: I: input activation, W: Weights, O: output activation. For Matrix Multiplication (MM):  A,B: operand tensors, Z: result tensor. Note that the designs have different dataflows, which are not listed. }
\label{tab:arch_summary}
\end{table*}

%% file: 04_overview.tex
\section{\tool~Overview}
\label{sec:high_level}
The design space taxonomy in Sec.~\ref{sec:classification} lays the foundation for the modeling methodologies of \tool, an analytical modeling framework that \emph{quantitatively} evaluates the processing speed and energy efficiency of sparse tensor accelerators. In this section, we will discuss the modeling challenges and \tool's key methodologies to address those challenges. 


\subsection{Modeling Challenges}\label{sec:challenges}
There are three key challenges associated with ensuring the modeling framework's speed, accuracy, and flexibility. 


\noindent\textbf{Multiplicative factors of the design space.}
To faithfully model various sparse tensor accelerator designs, the analysis framework needs to understand the compound impact of  their sparsity-specific design aspects (\eg, the diverse SAFs shown in Table~\ref{tab:arch_summary}) together with general design aspects (\eg, architecture topology, dataflow, etc.). Simultaneously modeling the interactions between a considerable number of design aspects incurs high complexity, slowing down the modeling process.
Building specific models for each design cannot scale to cover the entire design space, either.

\noindent\textbf{Tradeoff between accuracy and modeling speed.}
High fidelity modeling requires time-consuming \dataDependent analysis. Since sparsity characteristics impact a sparse accelerator's performance, carefully examining the exact data in each tensor could ensure accuracy. However, the downside of actual-data based analysis is that it can cause intolerable slowdown during mapspace exploration, especially for workloads with numerous and evolving data sets, \eg, DNNs. 

\noindent\textbf{Evolving designs/workloads.}
Finally, diverse and constantly evolving designs/workloads require flexibility and extendability in the modeling framework. Since the interactions between the processing schedules and workload data characteristics are convoluted,
the framework must be flexible and modularized enough to allow easy extensions for future designs/workloads.

\subsection{\tool~Solutions to the Challenges}
To solve the challenges, \tool~makes two important observations for sparse accelerator modeling: (1) the runtime behaviors of sparse accelerators (\eg, number of storage accesses and computes) can be progressively modeled; (2) the \dataDependent behavior in sparse accelerators can be statistically modeled with negligible errors.

\begin{figure}[tb]
	\centering
	\includegraphics[width=0.7\linewidth]{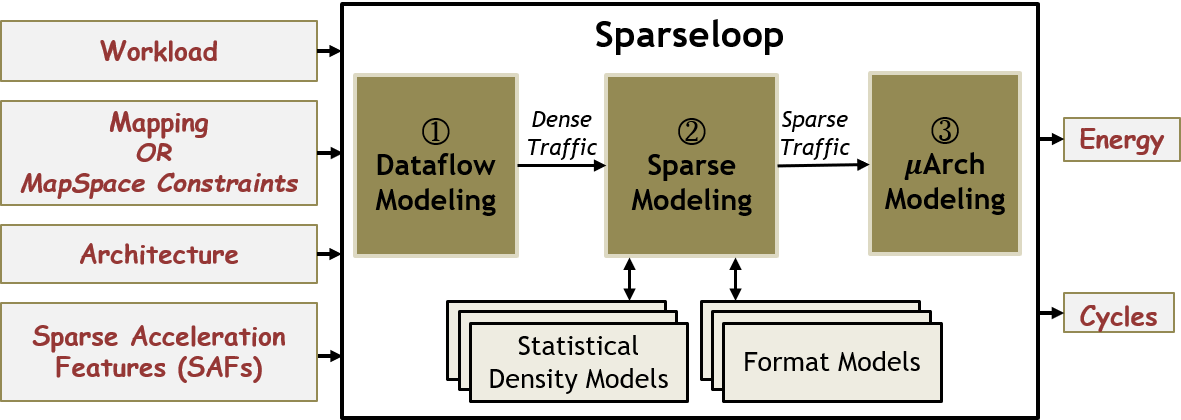}
	\caption{\tool~High-Level Framework. }
	\label{fig:high_level_infra}
\end{figure}

Based on observation (1), to maintain modeling complexity, \tool~performs decoupled modeling of distinct design aspects: \tool~evaluates dataflow independent of SAFs, as the storage accesses and computes introduced by the dataflow are irrelevant to how the IneffOps get eliminated; \tool~evaluates SAFs independent of micro-architecture, as the number of eliminated IneffOps introduced by the SAFs is orthogonal to the cost of performing each elimination or the savings brought by each eliminated IneffOp.
Thus, as Fig.~\ref{fig:high_level_infra} shows, \tool's modeling process is split into three steps, each with tractable complexity.
\begin{itemize}
    \item \emph{\textbf{Dataflow modeling}:} analyzes the uncompressed data movement and dense compute, \ie, dense traffic, incurred by the user-specified mapping input.
    \item \emph{\textbf{Sparse modeling}:} analyzes and reflects the impact of SAFs by filtering the dense traffic to produce sparse data movement and sparse compute, \ie, sparse traffic.
    \item \emph{\textbf{Micro-architecture modeling}:} analyzes the exact hardware operation cost (\eg, multi-word storage access cost) and generates the final energy consumption and processing speed based on the sparse traffic.
\end{itemize}

Based on observation (2), \tool~enables systematic recognition of the impact of SAFs at the sparse modeling step. To balance accuracy and speed, sparse modeling performs analysis based on \emph{statistical characterizations} of nonzero value locations in workload tensors and their sub-tensors, by leveraging various statistical density models.

Finally, as shown in Fig.~\ref{fig:high_level_infra}, to ensure extendability, the sparse modeling step interacts with statistical density models and per-dimension format models as decoupled modules so that these models can be extended to support future sparse workloads and representation formats.

%% file: 05_0_inputs.tex
\section{\tool~Framework}

We first discuss the inputs to \tool~in Sec.~\ref{sec:inputs}, and describe the  modeling steps in Sec.~\ref{sec:dataflow}, ~\ref{sec:sparse_modeling}, and ~\ref{sec:microarch}.

\subsection{Inputs}
\label{sec:inputs}

\begin{figure}[tb]
	\centering
	\includegraphics[width=0.7\linewidth]{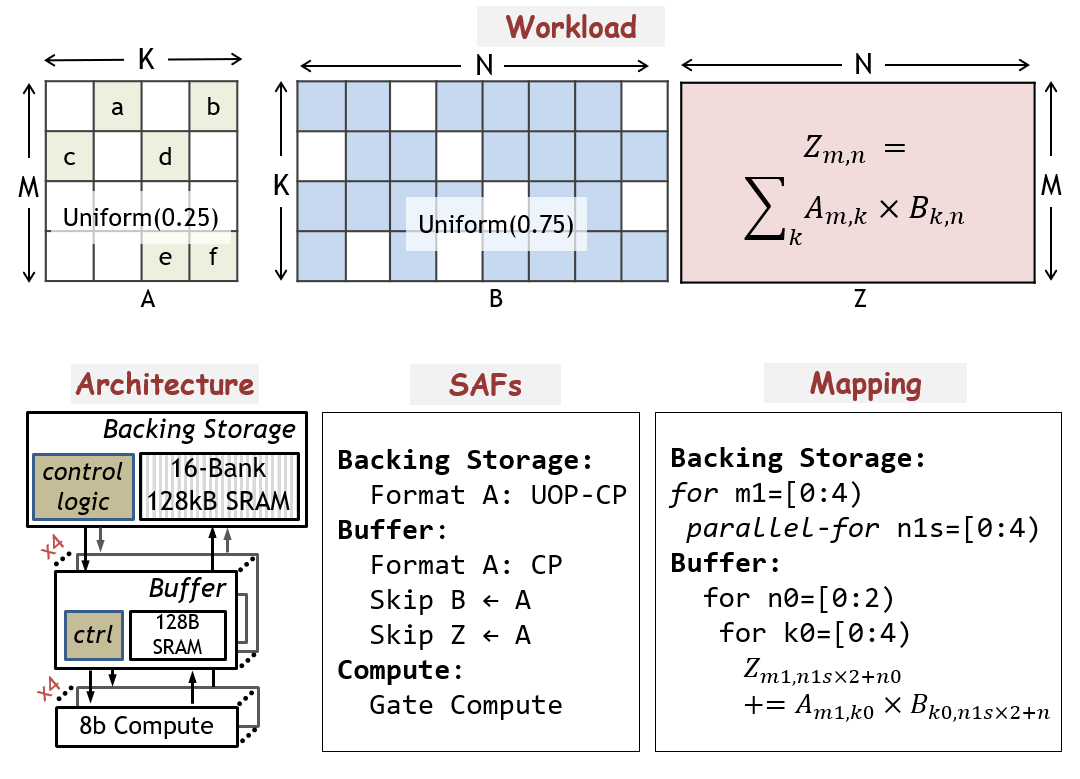}
	\caption{Example input specifications to \tool. Blank spaces in the workload tensors refer to locations with zeros. The locations of zeros are just for illustrative purposes.}
	\label{fig:running-example}
\end{figure} 

As shown in Fig.~\ref{fig:high_level_infra}, 
\tool~needs four inputs: workload specification, architecture specification, SAFs specification, and mapping or mapspace constraints. Fig.~\ref{fig:running-example} shows a set of example specifications to show the input semantics, and more detailed syntax can be found at~\cite{timeloop-github}.

\emph{Workload specification} describes the shape and statistical density characteristics of the workload tensors (\eg, in Fig.~\ref{fig:running-example}, A is 4x4 and has a density of 25\% with a uniform distribution). Workload specification also includes the tensor algorithm specification, which is based on the well-known Einsum notation~\cite{einsum,taco} (\eg, the matrix multiplication kernel is specified as $Z_{m,n} = \sum_{k} A_{m,k} \times B_{k,n}$, where the $A$ and $B$ values along the same $k$ dimension are reduced and the $m$ and $n$ dimensions are populated to the output tensor $Z$). \tool~ understands any algorithm described with an extended Einsum notation, similar to existing works~\cite{timeloop-ispass, extensor}.

\emph{Architecture specification} describes the hardware organization of the architecture (\eg, two levels of storage and four compute units) and the hardware attributes of the component in the architecture (\eg, \emph{Backing Storage} is 128kB).

\emph{SAFs specification} describes the SAFs applied to the storage or compute levels and the relevant attributes associated with each SAF (\eg, Fig.~\ref{fig:running-example} specifies skipping at \textit{Buffer}, with A as the leader and B as the follower). 

\emph{Mapping} describes an exact schedule for processing the workload on the architecture. It is represented by a set of loops~\cite{timeloop-ispass}. Each iteration of the \emph{for} loop represents a time step, and the iterations in a \emph{parallel-for} loop represent operations happen simultaneously at different spatial instances (\eg, $n1s$ loop shows that different columns of B are simultaneously processed in four \emph{Buffer}s). 

\emph{Mapspace Constraints} describes a set of constraints on allowed schedule (\eg, allowed loop orders). \tool~ then explores the potential mappings that satisfy the provided partial loops and locates the best one for a specific workload.

%% file: 05_1_dataflow_modeling.tex
\subsection{Step One: Dataflow Modeling}
\label{sec:dataflow}

Dataflow modeling derives the uncompressed data movement and dense compute, which we refer to as the \emph{dense traffic}. Such dense analysis has been studied in several existing works~\cite{timeloop-ispass, maestro, cosa, Zigzag, scale-sim}. Since each modeling step in \tool~is well-abstracted, various strategies can be plugged into \tool's modeling process. In our implementation, we adopt Timeloop's~\cite{timeloop-ispass} strategy.

\begin{figure}[tb]
	\centering
	\includegraphics[width=0.7\linewidth]{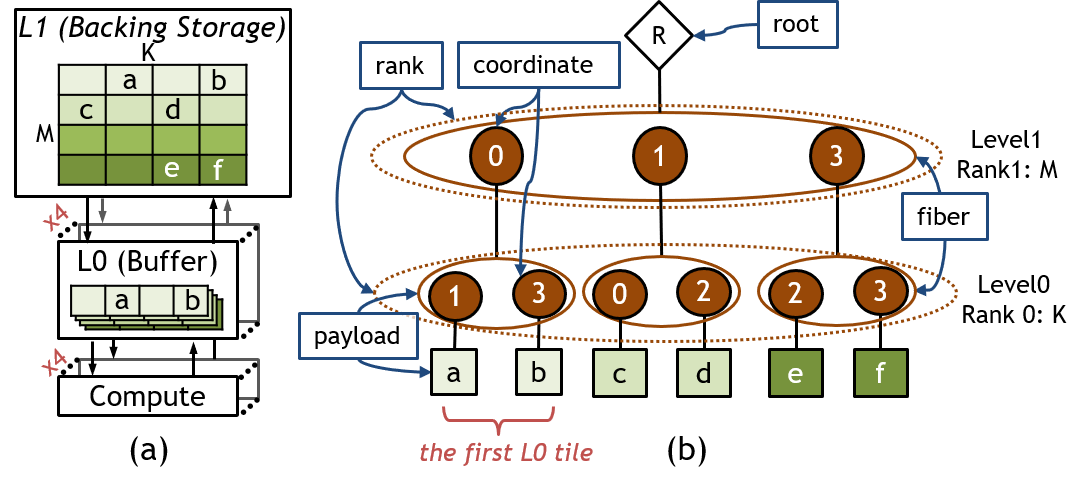}
	\caption{(a) Example coordinate-space tiling for tensor A based on inputs specified in Fig.~\ref{fig:running-example}. The shades represent tiles processed at different time steps. (b) Fiber tree representation of the tensor A. Each level of the tree corresponds to a \emph{rank} of the tensor and contains one or more \emph{fibers} that correspond to the rows or columns of the tensor. The leaves of the tree are the (nonzero) data values of the tensor.}
	\label{fig:tiling}
\end{figure}

Dataflow modeling is performed based on an abstract architecture topology (\eg, Fig.~\ref{fig:tiling}a shows the abstract representation of the architecture in Fig.~\ref{fig:running-example}), workload tensors' shapes, and the specified mapping. According to the mapping, each workload tensor is hierarchically partitioned into smaller tiles based on coordinates, with each tile stored in a specific storage level, and this process is referred to as \emph{coordinate-space tiling}~\cite{dnn_syn_lec}. For example, in Fig.~\ref{fig:tiling}a, at \emph{L1}, the tensor A in Fig.~\ref{fig:running-example} is partitioned into four tiles (with different shades of blue) based on the \emph{m1 for loop} in the mapping, each of which is a row of the tensor. Each tile is then sequentially sent to \emph{L0}. 
To derive the data movement for each storage level, dataflow modeling analyzes the stationarity of the tiles and the amount of data transferred, both temporally and spatially, between consecutive tiles. The number of computes is derived based on the input tensor algorithm. More detailed description of dense traffic calculations can be found in Timeloop~\cite{timeloop-ispass}.

%% file: 05_2_sparse_modeling.tex
\subsection{Step Two: Sparse Modeling}
\label{sec:sparse_modeling}
\begin{figure}[tb]
	\centering
	\includegraphics[width=0.9\linewidth]{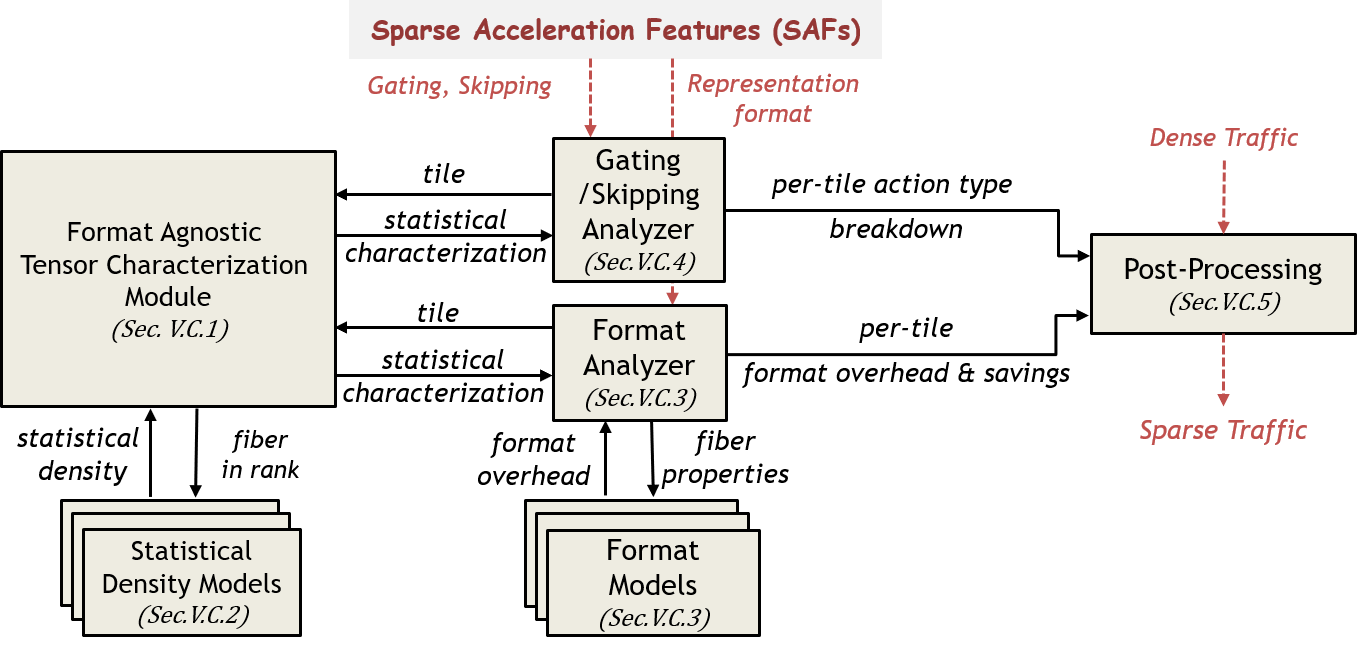}
	
	\caption{Various modules in sparse modeling step. Hashed red arrows refer to inputs and outputs of this step. The modules are labeled with their corresponding section numbers.}
	\label{fig:zoomed_in_sparse_modeling}
	
\end{figure}

The sparse modeling step is responsible for reflecting the overhead and savings introduced by various SAFs. As shown in Fig.~\ref{fig:zoomed_in_sparse_modeling}, sparse modeling first evaluates the impact of SAFs locally on per-tile traffic with SAF-specific analyzers, \ie, the \textit{Gating/Skipping Analyzer} and the \textit{Format Analyzer}, and then post processes the local traffic with simple scaling to reflect SAFs' impact on overall traffic. 

Such decomposition of local and global traffic analysis allows sparse modeling to reflect SAFs' impact \emph{on top of} the dense traffic to produce \emph{sparse traffic} for storage and compute units. We now discuss how each module in Fig.~\ref{fig:zoomed_in_sparse_modeling} interacts with others and the insight behind this design.

\subsubsection{\textbf{Format-Agnostic Tensor Description}} 

 As shown in Fig.~\ref{fig:zoomed_in_sparse_modeling},
  to allow tractable complexity and extendability, sparse modeling performs decoupled analysis of SAFs with different analyzers. Describing sparsity characteristics independent of representation format is core to performing such decoupled analysis. We adopt the \emph{fibertree} concept~\cite{dnn_syn_lec} to achieve format-agnostic tensor description. In Fig.~\ref{fig:tiling}b, we present the fibertree representation of the sparse tensor A stored in \textit{L1} of Fig.~\ref{fig:tiling}a.
 With the example, we first introduce the key fibertree concepts relevant to \tool. 

In fibertree terminology, each dimension of a tensor is called a \emph{rank}\footnote{A rank can also correspond to split or flattened dimensions.} and is named. Thus this 2D tensor has 2 ranks, with the rows being named $M$ (rank1), and columns being named $K$ (rank0).  In Fig.~\ref{fig:tiling}b, each level of the tree corresponds to a tensor rank in a specific order.  
Each rank contains one or more \emph{fibers}, representing the rows or columns of the tensor. Each fiber contains a set of coordinates and their associated \emph{payloads}. For intermediate ranks, the payload is a fiber from a lower rank (\eg, coordinate 0 in rank1 has a fiber in rank0 as its payload); for the lowest rank, the payload is a simple value. By omitting the coordinate for all-zero payloads, \ie, empty elements, a fibertree-based description accurately reflects the tensor's sparsity characteristics (\eg, rank1's fiber having empty coordinate 2 indicates that the third row is all-zero). 

 Each fiber in the tree corresponds to a \emph{tile} being processed. For example, in Fig.~\ref{fig:tiling}, the first tile processed in \textit{L0} corresponds to the first fiber in \emph{Rank K}. Thus, fibertree-based description enables format-agnostic \dataDependent analysis: to analyze the tiles of interest, the analyzers can examine the appropriate fibers to obtain sparsity information independent of the tensor's representation format.

\subsubsection{\textbf{Statistical Density Models}} 
\label{sec:data_density}

\begin{figure}[tb]
	\centering
	\includegraphics[width=0.6\linewidth]{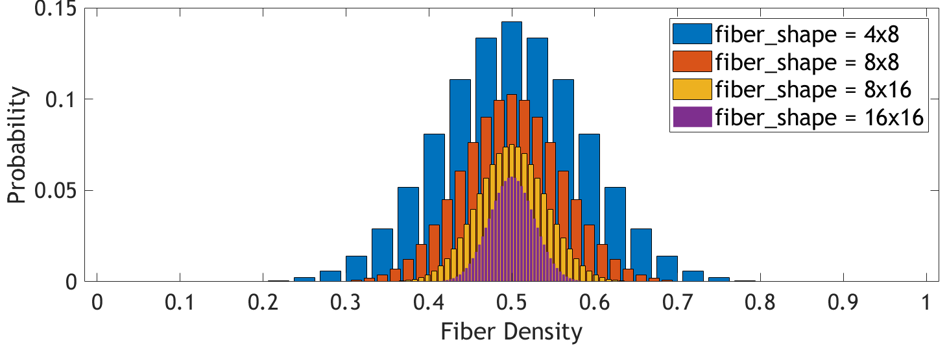}
	\caption{Fiber density probabilities for fibers with various shapes in a tensor with 50\% randomly distributed nonzeros.}
	\label{fig:pdf}
\end{figure}

Examining every fiber (thus analyzing the behavior of every tile) is too time-consuming for mapspace and design space exploration. To enable faster analysis, \tool~performs statistical characterizations of the fibers in the fibertree. As shown in Fig.~\ref{fig:zoomed_in_sparse_modeling}, \tool~can use various statistical density models of the workload tensor to derive statistical density for fibers in each rank (\eg, for the example in Fig.~\ref{fig:tiling}b, the fibers in rank0 have a density of 50\% with a probability of 0.75 and a density of 0\% with a probability of 0.25). For a given density model, the derived statistical density can differ significantly across fibers with different shapes (\ie, fibers from different ranks in the tree). 
For example, Fig.~\ref{fig:pdf} shows the distribution of fiber densities in a tensor with uniformly distributed non-zero values. In a uniform distribution, a tile's shape varies inversely with the deviation in its density. 


\input{tables/density_models} 
To estimate the density of fibers with a given shape, a density model either performs \emph{coordinate-independent} modeling (\ie, fibers at different coordinates have similar density distributions) or \emph{coordinate-dependent} modeling (\ie, fiber's density is a function of its coordinates). \tool~supports four popular density models: \emph{fixed-structured, uniform, banded, and actual data}. Table~\ref{tab:density_model} describes their properties and use cases in terms of relevant applications (\eg, randomly pruned DNNs~\cite{dual-side-tensor-core} and scientific simulations~\cite{banded-matrix-application}). The modularized implementation of density models ensures \tool's extensibility to modeling of emerging workloads with different nonzero value distributions.

\subsubsection{\textbf{Format Analyzer}} 
\label{sec:format}
With fibers statistically characterized, the format analyzer is responsible for deriving the representation overhead for the tiles stored in different storage levels. Since different tiles correspond to different fibers, it's important for the analyzer to identify the tile in each storage and obtain the appropriate statistical characterization of the corresponding fiber from the \emph{format-agnostic tensor characterization module}.

As shown in Fig.~\ref{fig:zoomed_in_sparse_modeling}, for each fiber, the analyzer statistically models the overhead of each rank with the appropriate per-rank \emph{Format Model}. Different formats introduce different amounts of overhead. For example, the \textit{RLE} format model calculates the overhead based on the number of non-empty elements in the fiber, $Overhead_{RLE}$ $=$ $\#\, non$-$empty$-$elements \times run$\_$length$\_$bitwidth$; whereas, the bitmask (\textit{B}) format model produces the same overhead regardless of fiber density, $Overhead_{B}$ $=$ $total \, \# \, elements \times 1$. The statistical format overhead allows \tool~to derive important analytical estimations, \eg, the average and worst-case overhead. 
\tool~supports five per-rank format models: \emph{B}, \emph{CP}, \emph{UOP}, \emph{RLE}, and \emph{Uncompressed B}, and thus supports any representation format that can be described with these models. The framework can be easily extended to support other formats.

\subsubsection{\textbf{Gating/Skipping Analyzer}}

The \emph{Gating/Skipping Analyzer} evaluates the amount of eliminated IneffOps introduced by each gating/skipping SAF. Since gating/skipping focuses on improving efficiency for each tile being transferred and/or each compute being performed, regardless of the total number of operations, the analyzer evaluates the impact of SAFs locally on per-tile traffic and breaks down the original per-tile dense traffic into three fine-grained action types: i) actually happened, ii) are skipped, and iii) are gated.

As discussed in Sec.~\ref{sec:classification}, gating/skipping is based on various intersections, which eliminate IneffOps by locating the empty tiles, \ie, tiles with all zeros. In a leader-follower intersection, when the leader tile is empty, the IneffOps associated with the follower are eliminated. Whereas in a double-sided intersection, any tile being empty leads to eliminations of IneffOps associated with the other tile. Since a double-sided intersection can be modeled as a pair of leader-follower intersections ($B \leftrightarrow A = B \leftarrow A + A \leftarrow B$), we focus on discussing the modeling of SAFs based on leader-follower intersections. 

\begin{figure}[tb]
	\centering
	\includegraphics[width=0.8\linewidth]{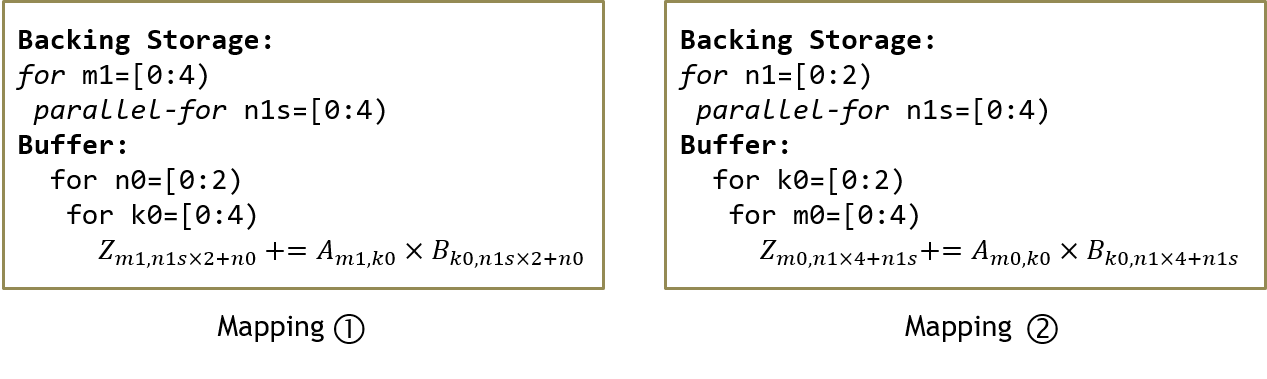}
	\caption{Example mappings that lead to intersections with different impact.}
	\label{fig:gs_analyzer}
\end{figure} 

The key to modeling the amount of eliminated IneffOps introduced by a SAF based on leader-follower intersection is to correctly identify the associated leader and follower tiles, and thus the fibers representing the tiles in the fibertree. Since different fibers can have significantly different probability of being empty, the same SAF can lead to different impacts. We observe that the leader and follower tiles of a specific intersection can be determined based on the data reuse defined in the mapping. For example, Fig.~\ref{fig:gs_analyzer} shows two mappings that lead to different intersection behaviors for  $Skip\>B\leftarrow A$ at \textit{Buffer}. In \emph{Mapping \circled{1}}, since the innermost $k0$ loop iterates through different pairs of A and B values, a specific $B_{k,n}$ is only used to compute with a single $A_{m,k}$. Thus, the leader tile is a single A value and the follower tile is a single B value, \ie, if $A_{m,k}$ is zero, the access to $B_{k,n}$ will be eliminated. Whereas in \emph{Mapping \circled{2}}, since the innermost $m0$ loop only iterates through different A values, a specific $B_{k,n}$ is reused across $A_{0:3,k}$ (\ie, a column of A). Thus, the leader tile is $A_{0:3,k}$ and the follower tile is a single B value $B_{k,n}$, \ie, if the entire column of A is empty, the access to $B_{k,n}$ will be eliminated. Since it is less likely for the entire column of A to be empty, under \emph{Mapping \circled{2}}, $Skip\>B\leftarrow A$ eliminates fewer IneffOps (\eg, columns of A are never empty in Fig.~\ref{fig:running-example}).

Based on the mapping and the statistical fibertree characterization, the analyzer defines the behaviors of each gating or skipping SAF, and derives a breakdown of the original dense traffic for each tile into fine-grained action types (\eg, for each B tile transferred from \emph{Buffer} to \emph{Compute}, there are 50\% skipped reads, 50\% actual reads, 0\% gated reads). Furthermore, when a gating/skipping SAF is applied to upper storage levels in the architecture, the analyzer propagates the savings introduced to lower levels (\eg, for the architecture in Fig.~\ref{fig:running-example}, skipping at \emph{Backing Storage} reduces operations happened at both \emph{Buffer} and \emph{Compute}). 

\subsubsection{\textbf{Traffic Post-processing}} 
As shown in Fig.~\ref{fig:zoomed_in_sparse_modeling}, after the analyzers evaluate the impact of their respective SAFs based on per-tile traffic, sparse modeling performs post-processing to first reflect the interactions between the SAFs (\eg, how much format overhead is skipped due to a skipping SAF) and then scale the per-tile breakdowns based on the number of tiles transferred to derive the final sparse traffic.

%% file: tables/density_models.tex
\begin{table}[tb]
\centering
\resizebox{\columnwidth}{!}{
\begin{tabular}{c||c|c}
\textbf{\begin{tabular}[c]{@{}c@{}}Density Models\end{tabular}} &
  \textbf{\begin{tabular}[c]{@{}c@{}}Sparsity Pattern\end{tabular}} &
  \textbf{\begin{tabular}[c]{@{}c@{}}Example Applications \end{tabular}} \\ \hline
  \rule{0pt}{15pt}
\textbf{\begin{tabular}[c]{@{}c@{}}Fixed structured \end{tabular}} &
  \begin{tabular}[c]{@{}c@{}}Even distribution, Coord. independent \end{tabular} &
  \begin{tabular}[l]{@{}c@{}}Structurally pruned  DNNs ~\cite{ampere-white-paper}\end{tabular} \\ \hline
 \rule{0pt}{16pt}
\textbf{Uniform} &
  \begin{tabular}[c]{@{}c@{}}Random distribution, Coord. independent\end{tabular} &
  \begin{tabular}[l]{@{}c@{}}Randomly pruned DNNs~\cite{dual-side-tensor-core} \& Activation sparsity\end{tabular} \\ \hline
 \rule{0pt}{12pt}
\textbf{Banded} &
  \begin{tabular}[c]{@{}c@{}}Diagonal distribution,  Coord. dependent\end{tabular} &
  \begin{tabular}[l]{@{}c@{}}SuiteSparse~\cite{suitesparse}, Scientific simulations~\cite{banded-matrix-application}\end{tabular} \\ \hline
\rule{0pt}{15pt}
\textbf{\begin{tabular}[c]{@{}c@{}}Actual Data\end{tabular}} &
  \begin{tabular}[c]{@{}c@{}}Non-statistical, Coord. dependent \end{tabular} &
   \begin{tabular}[l]{@{}c@{}}Graph analytics with special patterns~\cite{social-networks-graph}\end{tabular} \\ \hline
   
\end{tabular}
}
\caption{Summary of density models supported by \tool. New models can be easily added via \tool's interface.}
\label{tab:density_model}
\end{table}

%% file: 05_3_microarch_modeling.tex
\subsection{Step Three: Micro-architectural Modeling}
\label{sec:microarch}

Micro-architecture modeling first evaluates the validity of the provided mapping. A mapping is valid only if the largest tiles, which are derived based on statistical tile densities and format overheads, meet the capacity requirement of their corresponding storage levels. 
If the mapping is valid, micro-architecture modeling evaluates the impact of micro-architecture on generated sparse traffic. The analysis focuses on capturing general micro-architectural characteristics, \eg, segmented block accesses for storage levels, instead of the design-specific micro-architectural analysis, \eg, impact of an exact routing protocol.

Micro-architectural modeling then evaluates the processing speed and energy consumption. For processing speed, cycles are spent for \emph{actual} and \emph{gated} storage accesses and computes. The model considers available bandwidth at each level in the architecture to account for bandwidth throttling. For energy consumption, we use an energy estimation back end (\eg, Accelergy~\cite{accelergy-iccad}) to evaluate the cost of each fine-grained action, which is combined with its corresponding sparse traffic to derive accurate energy consumption. 


%% file: 06_1_speed.tex
\section{Evaluations}
We first introduce our experimental methodology and then demonstrate that \tool~is fast and accurate.
\subsection{Methodology}
\tool~ is implemented in C++ on top of Timeloop\cite{timeloop-ispass}, an analytical modeling framework for \emph{dense tensor accelerators}. \tool~reuses Timeloop's dataflow analysis, adds the new sparse modeling step, and improves Timeloop's micro-architectural analysis to account for the impact of various fine-grained actions introduced by the SAFs.  \emph{As a result, \tool~allows modeling of both dense and sparse tensor accelerators in one unified infrastructure}. We use Accelergy~\cite{accelergy-iccad, accelergy-ispass} as the energy estimation back end. 
For DNN workloads, \tool~performs per-layer evaluations \emph{with the appropriate dataflow and SAFs}, and aggregates the results to derive the energy/latency for the full network. This methodology is consistent with state-of-the-art tensor accelerator modeling frameworks~\cite{timeloop-ispass, cosa, maestro, Zigzag}. Experimental results in the next sections are all evaluated on an Intel Xeon Gold 6252 CPU. 

\subsection{Simulation Speed}
\label{sec:exp_speed}
Fast modeling speed allows designers to quickly explore each design's large mapspace as well as various designs. We evaluate \tool's modeling speed with the metric \emph{computes simulated per host cycle} (CPHC), which refers to the number of accelerator computes simulated for each cycle in the host machine that runs the modeling framework. 
CPHC carries similar information as MIPS (million instructions per second), a popular metric for evaluating simulators for conventional processors. 

Detailed cycle-level simulators often have CPHCs that are lower than 1. For example, STONNE~\cite{STONNE} has CHPCs that are less than 0.5 when running popular DNN layers with various architecture configurations, \eg, number of rows and columns in the compute array. The main reasons include: i) instead of statistical analysis, cycle-level simulators iterate through actual data to perform all computations, which take significant time for large workloads with millions of computations such as DNNs; ii) detailed control logic needs to be simulated for every cycle and all the components (\eg, memory interface protocols, exact intersection checks).

\tool~achieves much higher CPHCs with its analytical modeling approach since \tool~avoids performing analysis on all computations by performing statistical analysis on transient and steady state design behaviors only; and does not simulate detailed cycle-level control logic. Table~\ref{tab:simulation_speed} shows \tool's CPHCs for example DNN accelerators~\cite{eyeriss-v1, eyeriss-v2, scnn} running representative workloads~\cite{ResNet, bert, VGG, alexnet}. The CPHCs are dependent on accelerator architecture characteristics (\eg, SAFs complexity, number of levels, etc.), employed dataflow, and DNN workload characteristics (\eg, sparsity, tensor shapes, number of layers, etc.). For example, compared to Eyeriss V2 and SCNN, Eyeriss' less powerful SAF support (more details in Table~\ref{tab:arch_summary}) always introduces lower SAF modeling complexity and more simulated computes, leading to a higher CPHC. Overall, \tool~is over $2000\times$ faster compared to STONNE~\cite{STONNE}. 
\input{tables/simulation_speed}


%% file: tables/simulation_speed.tex


\begin{table}[tb]
\centering
\resizebox{0.5\columnwidth}{!}{
\begin{tabular}{c|cccc}
\multirow{2}{*}{\textbf{\begin{tabular}[c]{@{}c@{}}Accelerator \\ Designs\end{tabular}}} & \multicolumn{3}{c}{\textbf{Workloads}} \\
                                                                       & ResNet50   & BERT-base  & VGG16     & AlexNet     \\ \hline \midrule 
Eyeriss                                                                & 5.2k       & 13.3k     & 53.8k     & 21.4k        \\
Eyeriss V2 PE                                                          & 2.7k       & 12.5k     & 20.4k     & 13.2k       \\
SCNN                                                                   & 1.1k       & 4.3k     & 3.7k      & 5.2k       
\end{tabular}
}
\caption{\emph{Computes simulated per host cycle (CPHC)} for designs modeled by \tool. Compared to cycle-level tensor accelerator simulator STONNE~\cite{STONNE}, which has less than 0.5 CPHC, \tool~is over 2000$\times$ faster.}
\label{tab:simulation_speed}
\end{table}

%% file: 06_2_validation.tex
\subsection{Validation}
\label{sec:exp_val}
High modeling accuracy, in terms of both absolute values and relative trends, allows designers to correctly analyze design trade-offs at an early stage. To demonstrate \tool's accuracy, we validate on five well-known accelerator designs: SCNN~\cite{scnn}, Eyeriss~\cite{eyeriss-v1}, Eyeriss V2~\cite{eyeriss-v2}, and dual-side sparse tensor core (DSTC)~\cite{dual-side-tensor-core}, and Sparse Tensor Core (STC)~\cite{ampere-white-paper}. \textbf{Overall, \tool~maintains relative trends and achieves 0.1\% to 8\% average error.} Based on available information from existing work, validations are performed on baseline models that capture increasing amount of design details: from analytical models based on statistical sparsity patterns to cycle-level models/real hardware designs based on actual sparsity patterns.  At a high-level, common sources of error include: 1) statistical approximation of actual data 2) approximated component characteristics 3) approximated impact of design-specific micro-architectural implementations. Table~\ref{tab:validation_summary} summarizes the validations.
\input{tables/validation_summary}

In the next sections, we present more detailed validation discussions. In order to validate our work against prior works, we need to use the workloads reported in those works, despite the reported workloads being old  (though popular at the time of the work's publication) or different across designs. This is mainly due to the fact that other workloads are either not available in the reported results or not directly supported by the available simulators. 

\begin{figure}[ht]
    
    \begin{minipage}[b]{0.3\linewidth}
    \centering
    \includegraphics[width=\textwidth]{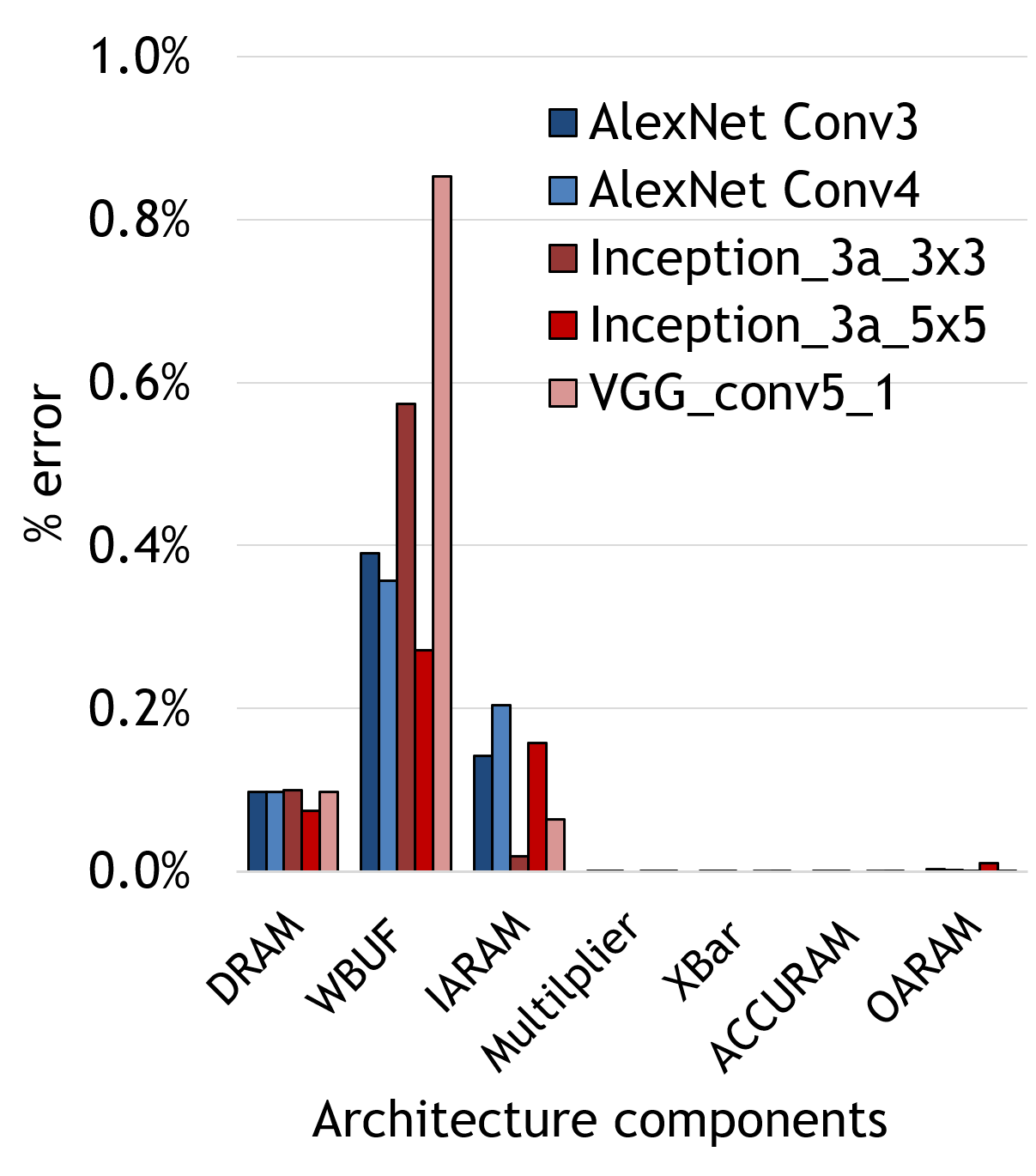}
    \caption{Runtime activity (\ie, storage access and compute count) validation for SCNN~\cite{scnn} . Achieves less than 1\% error for all storage and compute components in the architecture.}
	\label{fig:scnn_validation}
    \end{minipage}
    \hspace{0.3cm}
    \begin{minipage}[b]{0.33\linewidth}
    \centering
    \includegraphics[width=\textwidth]{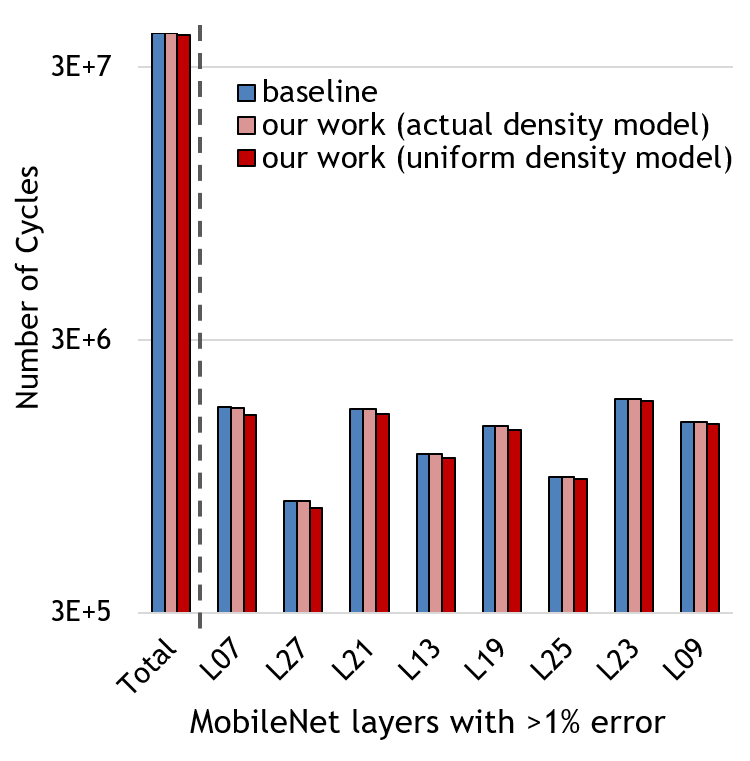}
    \caption{Processing latency validation for Eyeriss V2 processing element~\cite{eyeriss-v2} running MobileNet~\cite{mobilenets}. For ease of presentation, we only show total cycle counts and layers with more than 1\% error.}
	\label{fig:eyeriss_v2_validation}
    \end{minipage}
    \hspace{0.3cm}
    \begin{minipage}[b]{0.3\linewidth}
    \centering
    \includegraphics[width=\linewidth]{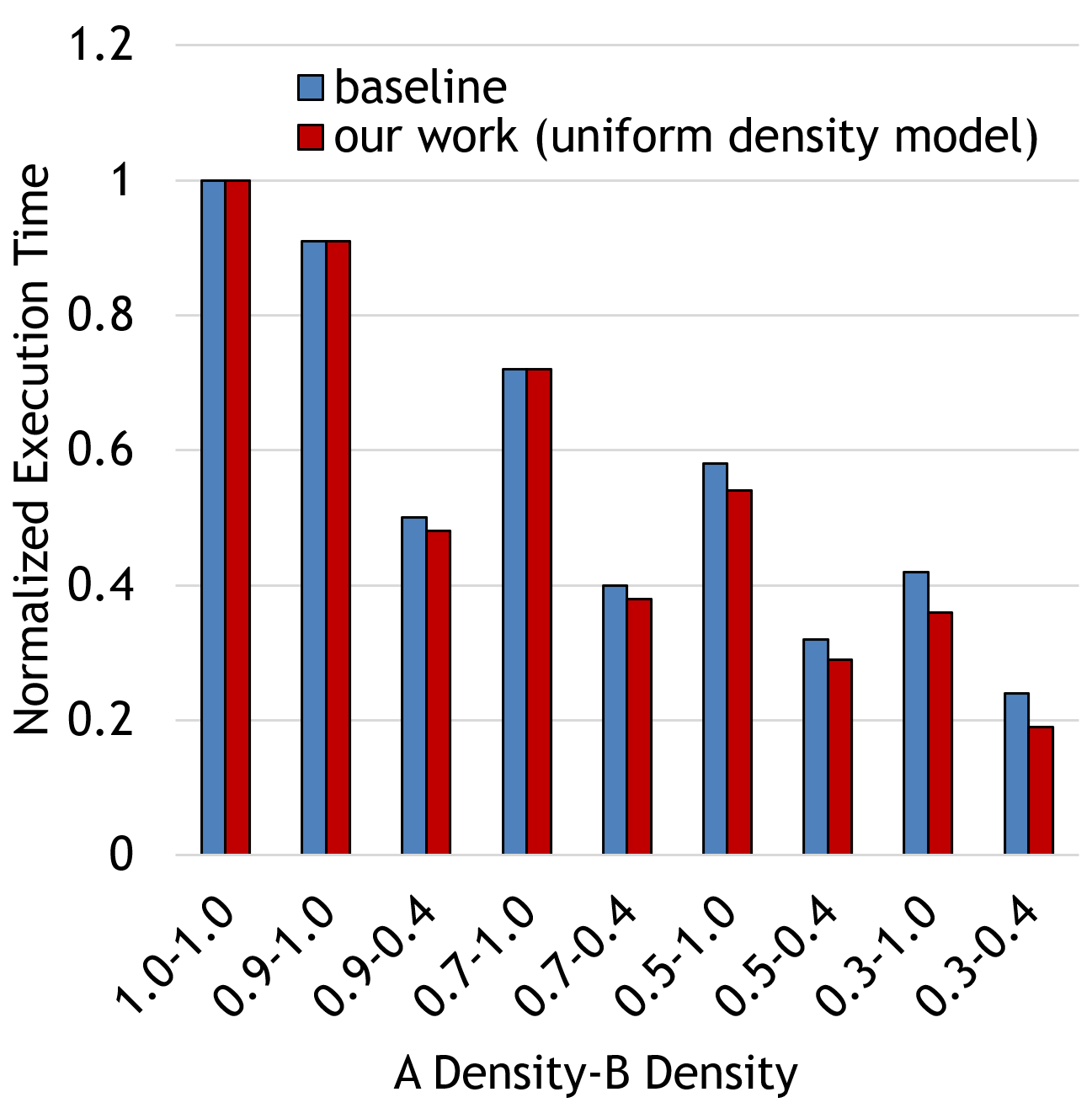}
	\caption{Processing latency of dual-side sparse tensor core~\cite{dual-side-tensor-core} running matrix multiplication workloads with various operand tensor densities, normalized to dense processing latency. Average error is 7.6\%.}
    \label{fig:dual-side-tensor-core}
    \end{minipage}
\end{figure}

\subsubsection{SCNN}
\label{sec:scnn}
We first validate \tool~on SCNN~\cite{scnn} with a customized simulator that was used in the paper: it performs analytical modeling based on \emph{statistically} characterized data. SCNN baseline assumes uniform distribution and captures the runtime activities of the components in the architecture (\eg, the number of reads and writes to various storage levels). 
Fig.~\ref{fig:scnn_validation} shows the error rate of the runtime activities for each component in the architecture. \tool~, running with a uniform density model, is able to capture all runtime activities accurately with less than 1\% error for all components in the architecture. 

\subsubsection{Eyeriss V2 }
\label{sec:eyeriss_v2}


Since Eyeriss V2's SAFs are implemented in its processing element (PE), we focus on validating the PE based on a baseline model that performs \emph{actual sparsity pattern} based analytical modeling. To quantitatively demonstrate the sources of error, we validate \tool~with both an actual-data density model and a uniform density model.

Fig.~\ref{fig:eyeriss_v2_validation} shows the validation on the number of cycle counts. In terms of total cycle counts for processing the entire MobileNet~\cite{mobilenets}, \tool~achieves more than 99\% accuracy and is able to capture the relative trends across different layers with both density models. However, for layers with both sparse operands compressed, modeling based on a uniform density model results in up to 7\% error for layer27. Fig.~\ref{fig:eyeriss_v2_validation} shows the layers with more than 1\% error. The errors are mainly attributed to the statistical approximation of the expected nonempty intersection ratio between two sparse tensors, as the exact nonempty ratio deviates from case to case, \eg, when both operands have identical nonzero value locations, the intersection nonempty ratio is equal to the tensor densities. With an actual-data density model, \tool~accounts for the exact intersections, and thus accurately captures the cycle counts (at the cost of a slower modeling speed). 

\subsubsection{Dual-Side Sparse Tensor Core}
\label{sec:dstc}

For DSTC, the baselines are also obtained directly from the papers whose reported results are based on a cycle-level simulator that is validated on real hardware~\cite{accelsim}.
We validate on the normalized processing latency running matrix multiplication workloads with various operand tensor density degrees, as shown in Fig.~\ref{fig:dual-side-tensor-core}. We modeled the tensors with a uniform density model, captured the performance trends across density degrees, and obtained an average error of less than 8\%. In addition to errors introduced by deviations from the expected nonempty intersection ratio, \tool~also performs optimistic modeling of micro-architectural details. More specifically, \tool~assumes no storage bank conflicts but DSTC's baseline results contain bank conflicts when operand tensors are relatively sparse (\eg, 30\% density), thus introducing higher processing latency. 

\subsubsection{Eyeriss}
\label{sec:eyeriss}
We validate on Eyeriss~\cite{eyeriss-v1} with baselines obtained from the paper and based on taped-out silicon.
We first validate DRAM compression rates for AlexNet~\cite{alexnet}, as shown in Table~\ref{tab:v1_compression}. Overall, we achieve 1\% error on average and the discrepancy could be due to imperfect compression with the actual data.
We also validate on the PE array energy reduction ratio due to on-chip gating. Eyeriss claims that the energy savings of the processing elements can achieve 45\%.  Our results show a max energy efficiency improvement of 43\%. The discrepancy could be due to not modeled PE components with unknown energy characteristics.
\input{tables/v1_compression}

\subsubsection{Sparse Tensor Core}
\label{sec:stc}
\begin{figure}[tb]
	\centering
	\includegraphics[width=0.7\linewidth]{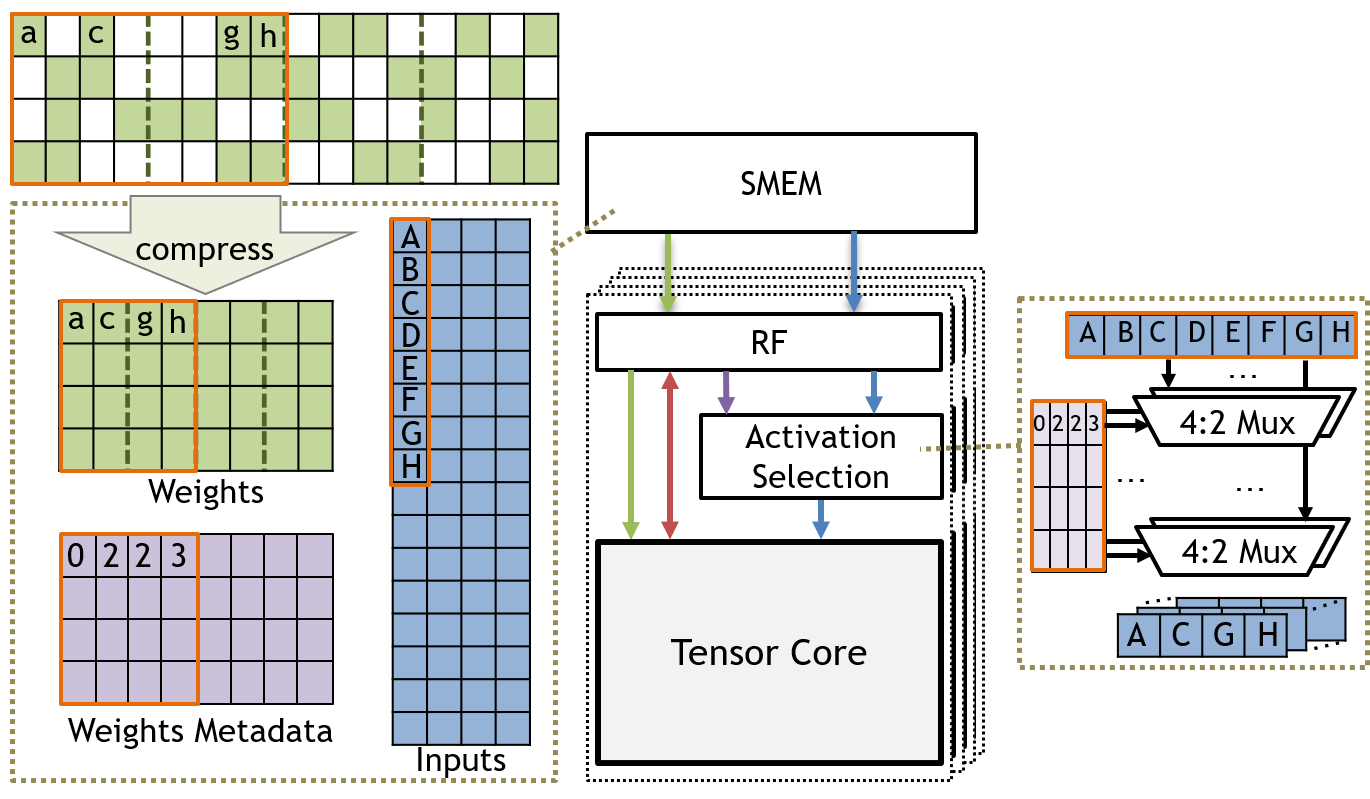}
	\caption{Modeled sparse tensor core architecture (including the \emph{SMEM} in streaming processor for a more holistic view) and example processing of a 2:4 structured workload.}
	\label{fig:stc-arch}
\end{figure}

Finally, we validate the Ampere GPU's sparse tensor core accelerator (STC) based on publicly available architecture descriptions~\cite{stc-isscc, dissecting-stc, ampere-white-paper}.
STC focuses on accelerating structured sparse workloads with a 2:4 sparsity structure, which demands at most two nonzero values in every block of four values. Fig.~\ref{fig:stc-arch} shows the high-level STC architecture and an example processing flow of a 2:4 structured sparse matrix multiplication workload (algorithm defined in Fig.~\ref{fig:running-example}). We will discuss more details about STC in Section~\ref{sec:stc-study}.

To validate STC, we use the fixed structured density model parameterized with the 2:4 structure along each channel to model the structured sparse weight tensor. Existing work reports that STC achieves 2$\times$ speedup compared to dense processing~\cite{ampere-white-paper, stc-isscc, dissecting-stc}. Because of the fully defined behaviors with the structured sparsity, \tool~also produces an exact 2$\times$ speedup (\textit{STC} design in Fig.~\ref{fig:stc-study}), achieving 100\% accuracy. 

%% file: tables/validation_summary.tex

\begin{table*}[tb]
\resizebox{\textwidth}{!}{

\begin{tabular}{c|cccc|c|l}
\toprule

\multirow{2}{*}{\textbf{\begin{tabular}[c]{@{}c@{}}Accelerator \\ Design\end{tabular}}} &
  \multicolumn{4}{c|}{\textbf{Baseline Model}} &
  \multirow{2}{*}{\textbf{\begin{tabular}[c]{@{}c@{}}Average \\ Accuracy\end{tabular}}} &
  \multicolumn{1}{c}{\multirow{2}{*}{\textbf{Major Sources of Error}}} \\ \cline{2-5} \rule{0pt}{15pt}
 &
  \multicolumn{1}{c|}{Source} &
  \multicolumn{1}{c|}{Type} &
  \multicolumn{1}{c|}{\begin{tabular}[c]{@{}c@{}}Sparsity \\ Pattern\end{tabular}} &
  Output &
   &
  \multicolumn{1}{c}{} \\ \hline 
 \rule{0pt}{10pt}
\textbf{SCNN} &
  \multicolumn{1}{c|}{\multirow{2}{*}{\begin{tabular}[c]{@{}c@{}}Simulators obtained \\ from authors~\cite{scnn, eyeriss-v2}\end{tabular}}} &
  \multicolumn{1}{c|}{\multirow{2}{*}{\begin{tabular}[c]{@{}c@{}}Design-specific \\ Analytical\end{tabular}}} &
  \multicolumn{1}{c|}{Statistical} &
  Runtime activities &
  99.9\% &
  None \\ \cline{1-1}  \cline{4-7}
  \rule{0pt}{10pt}
\textbf{Eyeriss V2 PE} &
  \multicolumn{1}{c|}{} &
  \multicolumn{1}{c|}{} &
  \multicolumn{1}{c|}{\multirow{3}{*}{\begin{tabular}[c]{@{}c@{}} \\ \\ Actual\end{tabular}}} &
  Processing latency &
  $>$98\% &
  Statistical approximations \\ \cline{1-3} \cline{5-7} 
  \rule{0pt}{18pt}
\textbf{Eyeriss} &
  \multicolumn{1}{c|}{\multirow{2}{*}{\begin{tabular}[c]{@{}c@{}} \vspace{-5pt} \\ Results directly \\ from paper \\ or technical report \\ \cite{eyeriss-v1,dual-side-tensor-core, ampere-white-paper}\end{tabular}}} &
  \multicolumn{1}{c|}{Real hardware} &
  \multicolumn{1}{c|}{} &
  \begin{tabular}[c]{@{}c@{}}Compression rate\\ Energy savings\end{tabular} & $>$95\% &
  \begin{tabular}[c]{@{}l@{}}(1) Statistical approximations\\ (2) Approximated component \\ energy characterizations\end{tabular} \\ \cline{1-1} \cline{3-3} \cline{5-7} 
  
\rule{0pt}{20pt}
\textbf{\begin{tabular}[c]{@{}c@{}}DSTC\end{tabular}} &
  \multicolumn{1}{c|}{} &
  \multicolumn{1}{c|}{\begin{tabular}[c]{@{}c@{}}Cycle-level simulator \\validated on \\ silicon~\cite{accelsim}\end{tabular}} &
  \multicolumn{1}{c|}{} &
  Processing latency &
  92.4\% &
  \begin{tabular}[c]{@{}l@{}} (1) Statistical approximations; \\ (2) Optimistic modeling \\ of microarchitectural details\end{tabular} \\ \cline{1-1} \cline{3-3} \cline{5-7} 
  
  \rule{0pt}{18pt}
  \textbf{\begin{tabular}[c]{@{}c@{}} STC\end{tabular}} &
  \multicolumn{1}{c|}{} &
  \multicolumn{1}{c|}{\multirow{1}{*}{\begin{tabular}[c]{@{}c@{}}Real hardware\end{tabular}}} &
  \multicolumn{1}{c|}{} &
  Processing latency &
  100\% &
  \begin{tabular}[c]{@{}l@{}} None  \\\textit{(structured sparsity introduces} \\ \textit{deterministic behaviors)} \end{tabular} \\ 
  
  \bottomrule
\end{tabular}
}
\caption{High-level summary of performed validations based on available data from existing work. Overall, \tool~achieves 0.1\% to 8\% average error across different designs. More details in Sections~\ref{sec:scnn}, \ref{sec:eyeriss_v2}, \ref{sec:dstc}, \ref{sec:eyeriss} and \ref{sec:stc}.}
\label{tab:validation_summary}
\end{table*}

%% file: tables/v1_compression.tex



\begin{table}[t]

\centering

\begin{tabular}{l|ccccc}
\multirow{1}{*}{}                        & Conv1     & Conv2     & Conv3     & Conv4     & Conv5     \\ \hline \hline \\[-1em]
\multicolumn{1}{c|}{\textbf{Eyeriss\cite{eyeriss-v1}}}    & 1.2   & 1.4   & 1.7   & 1.8   & 1.9   \\
\multicolumn{1}{c|}{\textbf{\tool}} & 1.2  & 1.4  & 1.7  & 1.9  & 1.9 
\end{tabular}

\caption{ Eyeriss\cite{eyeriss-v1} DRAM compression rate validation.}
\label{tab:v1_compression}
\end{table}

%% file: 06_4_stc_explore.tex
\section{Case Studies}
\label{sec:case-study}
In this section, we demonstrate \tool's flexibility with two case studies.

\subsection{Investigating Next Generation Sparse Tensor Core}
\label{sec:stc-study}

In recent years, various techniques have been proposed to add sparsity support to tensor core (TC). In this case study, we use \tool~to first compare two variations: the commercialized NVIDIA STC~\cite{ampere-white-paper} and a research-based proposal DSTC~\cite{dual-side-tensor-core}. Based on the comparison, we then discuss the potential opportunities for next-generation STC, and showcase an example design flow that uses \tool~to identify current STC design's limitations and explore various solutions to such limitations to unlock more potential.

\subsubsection{DSTC vs. NVIDIA STC}
\label{sec:dsctc_vs_stc}
We perform an apples-to-apples comparison of the two designs. Since both designs are TC-based, both architectures contain the \emph{SMEM-RF-Compute} hierarchy as shown in Fig.~\ref{fig:stc-arch}, and are controlled on allocated hardware resources, including compute, storage capacity, and memory bandwidth. To model realistic systems, we only provision a subset of a real GPU's  \emph{SMEM} bandwidth to the accelerators, since other processes running on the GPU share the same \emph{SMEM} storage.
At a high-level, DSTC employs complex sparsity support and a special outer product dataflow to exploit \textbf{arbitrary sparsity in both operands} to perform compression and skipping. In contrast, STC ignores input sparsity and uses low-overhead sparsity support to compress and perform skipping \textbf{on weights with 2:4 structured sparsity only}.

\begin{figure}[tb]
     \centering
     \begin{subfigure}[b]{0.7\linewidth}
         \centering
         \includegraphics[width=\linewidth]{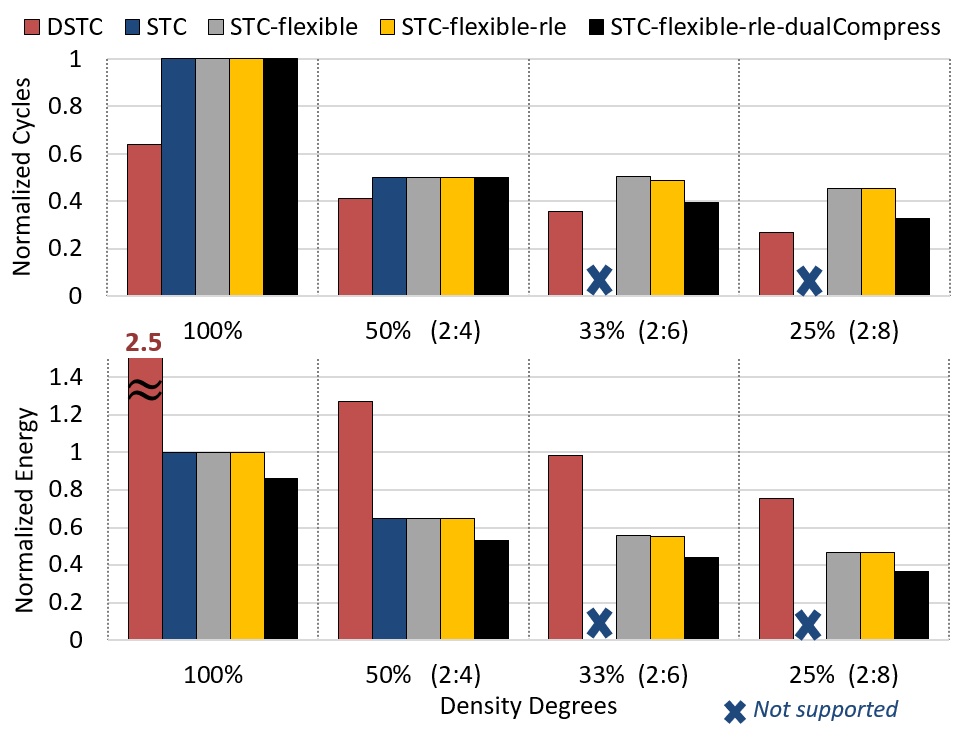}
     \end{subfigure}
        \caption{\tool's analysis on the  normalized total cycles spent and  energy-delay product for various designs of tensor core accelerator running representative ResNet50~\cite{ResNet} layers pruned to various sparsity degrees. The accelerators are controlled to have similar amount of hardware resources.}
        \label{fig:stc-study}
\end{figure}

Fig.~\ref{fig:stc-study} compares the cycles spent and energy consumed by DSTC and STC running ResNet50~\cite{ResNet} pruned to various sparsity degrees. ResNet50 contains sparse weights (if pruned) and sparse inputs. Compared to STC, DSTC's dataflow for supporting arbitrary sparsity incurs a significant amount of data movement. As a result, when processing denser workloads (\eg, unpruned ResNet50 in this example or BERT-like networks with dense input activations), even if DSTC is able to always introduce lower cycles counts, the savings brought by SAFs cannot compensate for the additional energy spent and thus the overall hardware efficiency is low. However, STC provides very limited support for different workloads. Furthermore, for sparser workloads, \eg, 25\% dense ResNet50 in Fig.~\ref{fig:stc-study}(a), despite DSTC's overhead, it's able to achieve a much higher overall efficiency because of the speedup introduced by a significant amount of skipping. 


\noindent{\textbf{Opportunities for STC:}}
\textbf{only supporting 2:4 sparsity in the current STC design leads to missed opportunities, as many modern DNNs can be pruned to $>$50\%  sparsity (structured~\cite{vector-sparse-tensor-core} or unstructured~\cite{deep-compression}) while maintaining reasonable accuracy. Thus, one possible feature for a next generation STC to have is to efficiently exploit the savings brought by more sparsity degrees but still keep the sparsity structured to reduce SAF overhead.}



\subsubsection{Naive STC Extension To Support More Ratios}

In order to extend the existing STC to support more sparsity degrees, we first introduce the existing high-level processing of STC running matrix multiplication workloads (algorithm defined in Fig.~\ref{fig:running-example}) with the default 2:4 structured sparsity. In the case of a DNN, tensor A in Fig.~\ref{fig:running-example} corresponds to the structured sparse weights in Fig.~\ref{fig:stc-arch}.

As shown in Fig.~\ref{fig:stc-arch}, the weight tensor is compressed with an offset-based coordinate payload format, where each nonzero carries an offset coordinate to indicate its position in the block of four values, \eg, the nonzero weight \textit{g} is the third element in its block, and thus carries a metadata of \textit{2}. This format matches our \emph{CP} format in earlier sections. The compressed weight tensor and the uncompressed input tensor are stored in \emph{SMEM}. For each iteration of processing, the weights, weight metadata, and dense inputs are fetched out. However, since inputs are uncompressed, as shown in Fig.~\ref{fig:stc-arch}, a\emph{ tile with four weights corresponds to a tile with eight inputs.} Thus, to ensure correctness, a 4:2 selection needs to be performed on the inputs for each block of four weights. Since only nonzero weights need to be processed, the 2:4 processing is $2\times$ faster than dense processing.

Thus, naively supporting more sparsity degrees in STC simply involves extending the above discussed sparsity support with input activation selection logic for more ratios, \eg, 2:6 and 2:8. We name this naive extension as \emph{STC-flexible}. As shown in Fig.~\ref{fig:stc-study}, \tool's modeling indicates that \emph{STC-flexible} does support and introduce extra energy reductions for lower density workloads. However, no desirable speedup is obtained with the higher sparsity, \eg, theoretically, 2:6 structured sparsity should introduce 3$\times$ speedup. \textbf{In fact, surprisingly, the \emph{baseline} processing barely brings any additional speedup with the naive extension for 2:6 and 2:8 workloads. }

\subsubsection{Identify Design Limitations}

\begin{figure}[tb]
	\centering
	\includegraphics[width=0.7\linewidth]{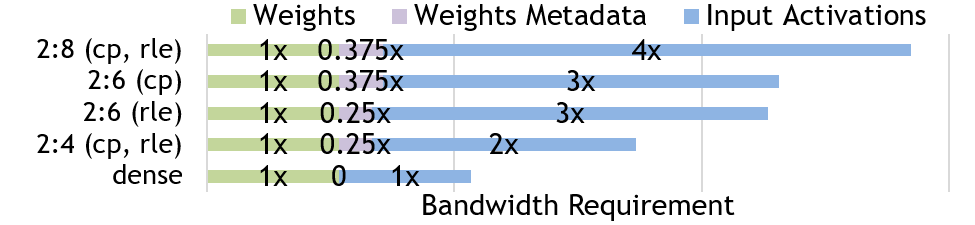}
	\caption{\tool's analysis on bandwidth requirements for getting ideal speedup for various operands and associated metadata (if any).}
	\label{fig:stc-bw-breakdown}
\end{figure}

\emph{STC-flexible}'s approach does not improve performance due to \emph{SMEM} bandwidth limitation.  Fig.~\ref{fig:stc-bw-breakdown} shows \tool's analysis on the required bandwidth for processing workloads with various sparsity ratios. To ensure full utilization of the tensor core, the same number of nonzero weights needs to be processed spatially regardless of the workload sparsity, \ie, we always need 1$\times$ weights as shown in Fig.~\ref{fig:stc-bw-breakdown}. As discussed above, STC stores inputs in uncompressed format. Thus, the sparser the weight tensor, the more inputs need to be fetched in a cycle, \eg, in Fig.~\ref{fig:stc-bw-breakdown}, 4$\times$ inputs need to be fetched for workloads with 2:8 sparse weights. In addition to the bandwidth pressure imposed by inputs, the metadata also needs to be described with more bits as the block size gets larger. The amount of additional metadata overhead is dependent on the chosen representation format, \eg, run length encoding (RLE) requires fewer bits than offset-based CP for 2:6 sparse workloads. \textbf{As a result, STC is bottlenecked by the limited bandwidth, which is provisioned for 2:4 structured sparsity, and thus cannot obtain the theoretical speedup for sparser workloads.}

\subsubsection{Explore Solutions to Overcome Limitations}

With \tool, we can perform early design stage exploration on potential solutions. Without loss of generality and for the ease of presentation, we discuss two example directions with low-hanging fruit: 1) improve representation format support to reduce metadata overhead; 2) introduce additional compression SAFs for inputs.

First, we evaluate if a different representation (compression) format can alleviate the overhead introduced by metadata, especially for 2:6 structured sparsity. Thus, as shown in Fig.~\ref{fig:stc-study}, we enabled \textit{RLE} support for \emph{STC-flexible} to form \emph{STC-flexible-rle}.
At a high-level, compared to the STC's original \emph{CP} support, \emph{RLE} support does provide similar or better processing speed. However, since the majority of the overhead comes from transferring actual data, the benefits are too insignificant to bring \emph{STC-flexible-rle} over DSTC. 

We then target the more important bottleneck: the uncompressed input data traffic. To solve the problem, we added bitmask-based compression to input such that both operands are compressed to form \emph{STC-flexible-rle-dualCompress} design in Fig.~\ref{fig:stc-study}. To keep the compute easily synced, we did not add input-based skipping. As a result, all of the obtained speedups come from bandwidth requirement reduction. As shown in Fig.~\ref{fig:stc-study},  \emph{STC-flexible-rle-dualCompress} can actually introduce similar speed even if it cannot exploit input sparsity for skipping. This is because even if DSTC exploits both operands for speedup, its dataflow has more frequent streaming of operands, introducing additional pressure to \emph{SMEM} bandwidth as well. \textbf{Thus, with this example, we have demonstrated that exploiting more sparsity does not guarantee more speedup, and it is very important to make sure the dataflow and SAF overhead is reasonable.} 

Overall, as shown in Fig.~\ref{fig:stc-study}, we derived \emph{STC-flexible-rle-dualCompress}  that, compared to DSTC, always introduces lower energy consumption and has similar processing speed most of the time for the studied sparsity degrees.

%% file: 06_3_case_study.tex
\subsection{Co-design of Dataflow, SAFs and Sparsity}
\label{sec:exp_case_studies}
Looking beyond the deep learning workloads and tensor core accelerators discussed in the previous case study, this section demonstrates how \tool \ can model workloads with more diverse sparsity degrees and accelerator designs that employ various dataflows and SAFs. 
With a set of small-scale experiments, we show various broad insights for designing sparse tensor accelerators: (1) the best  design for one application domain might not be the best for another; (2) combining more energy or latency saving features together does not always make the design more efficient. Thus, careful  co-design of dataflow, SAFs and sparsity is necessary for achieving desired latency/energy savings.

\subsubsection{Design Choices}
\noindent \textbf{Workloads}: We use matrix multiplication with sparse input tensors (spMspM) of various density degrees as example workloads. spMspM, represented as $Z_{m,n} = \sum_{k} A_{m,k} \times B_{k,n}$ as an Einsum, is an important kernel in many popular applications, such as scientific simulations, graph algorithms and DNNs, each of which can have different tensor density degrees. 

\noindent\textbf{Dataflows}: Given a hardware budget of 256 compute units and 128KB on-chip storage, we consider two  choices shown in Table~\ref{tab:case_study_archs}(a): (1) \emph{ReuseABZ} that reuses all tensors on-chip; (2) \emph{ReuseAZ} that doesn't have on-chip reuse for B. 

\noindent\textbf{SAFs}: As shown in Table~\ref{tab:case_study_archs}(b), we consider two sets of SAFs choices: (1) \textit{InnermostSkip} that performs $SkipB\leftrightarrow A$ at the innermost on-chip storage 
(2) \textit{HierarchicalSkip} that hierarchically performs $SkipB\leftrightarrow A$ at DRAM and innermost storage to reduce both off-chip and on-chip data movement.

\input{tables/case_study_archs}

\begin{figure}[tb]
	\centering
	\includegraphics[width=0.9\linewidth]{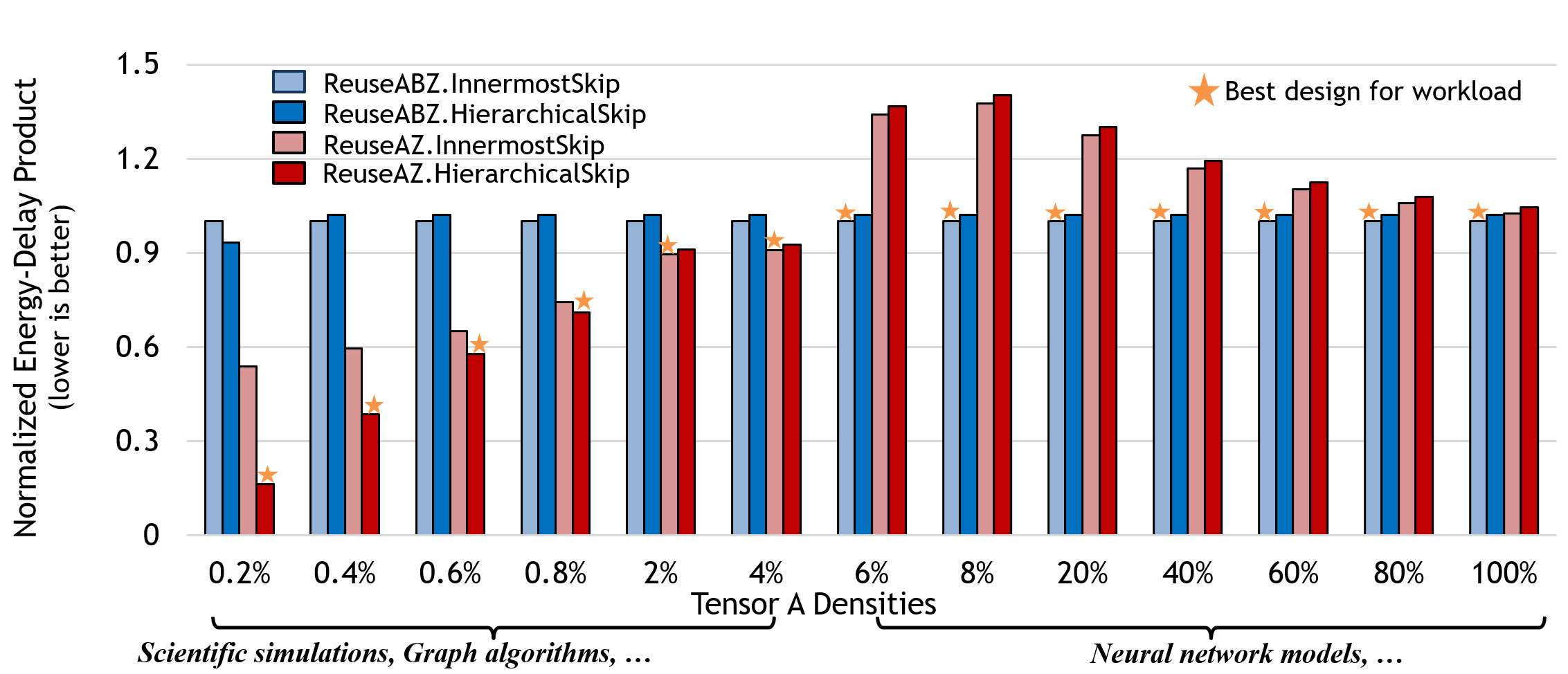}
	\caption{Normalized energy-delay product of different combinations of dataflow-SAFs running matrix multiplications with various density degrees, which are labeled with relevant example workloads. \tool~shows (1) dataflow and SAFs should be co-designed to ensure potential savings; (2) the correct combination needs to be chosen for different applications to realize the potential savings.}
	\label{fig:case_study_uniform}
\end{figure}

\subsubsection{Interactions Among Design Choices}
Fig.~\ref{fig:case_study_uniform} compares the energy-delay-product (EDP) of different dataflow-SAF combinations running spMspM with various A tensor density degrees. At each density degree, the EDPs are normalized to \emph{ReuseABZ.InnermostSkip}'s EDP. 

\emph{We first make the observation that the best design for one application domain might not be the best for another.} For example, while \textit{ReuseABZ.InnermostSkip} is the best design for NN workloads (\ie, A density $>$6\%), for sparser workloads, such as scientific simulations or graph algorithm,  this design is sub-optimal due to its large off-chip bandwidth requirement. On the other hand, \textit{ReuseAZ.HierarchicalSkip} performs the best with hyper-sparse workloads since this design performs early off-chip traffic eliminations, but it fails to reduce EDP with NN workloads due to its inability to perform effective off-chip intersections on denser operands and its lack of on-chip B reuse.  Thus, a design's dataflow-SAFs combinations need to be chosen based on target application's sparsity characteristics to realize potential savings.


\emph{We also show that combining more energy or latency saving features together does not always make the design more efficient.} For example, \emph{ReuseABZ.HierarchicalSkip} combines a dataflow that reuses all tensors with SAFs that skip both off-chip and on-chip traffic to form a design with the most number of latency/energy savings features. However, as shown in Fig.~\ref{fig:case_study_uniform}, \emph{ReuseABZ.HierarchicalSkip} is never the best design in terms of EDP.
This is because the \emph{ReuseABZ} dataflow prevents the off-chip skipping SAF from eliminating B's off-chip data movement. More specifically, since \emph{ReuseABZ} reuses each on-chip B tile for multiple A tiles, B tile transfers can be eliminated by the off-chip skipping SAF only when all values in its corresponding A tiles are zeros, which rarely happens. \textbf{Thus, dataflow and SAFs need to be carefully co-designed to ensure there exist opportunities for reasonable savings.}

%% file: tables/case_study_archs.tex
\begin{table}[tb]

\begin{tabular}{cccc}
\multicolumn{1}{l}{\textbf{(a)}}                                 & \multicolumn{1}{l}{}          & \multicolumn{1}{l}{}   & \multicolumn{1}{l}{}   \\ 
\multicolumn{1}{c|}{\multirow{2}{*}{\textbf{Dataflow Choices}}} & \multicolumn{3}{c}{\textbf{Tensor Reuse}}                                               \\ \cline{2-4} 
\multicolumn{1}{c|}{}                                  & \multicolumn{1}{c|}{A}        & \multicolumn{1}{c|}{B} & \multicolumn{1}{c}{Z} \\ \hline
\multicolumn{1}{c|}{\textbf{ReuseABZ}} & \multicolumn{1}{c|}{Innermost storage} & \multicolumn{1}{c|}{Shared buffer} & \multicolumn{1}{c}{Innermost storage} \\ \hline
\multicolumn{1}{c|}{\textbf{ReuseAZ}}  & \multicolumn{1}{c|}{Innermost storage} & \multicolumn{1}{c|}{None}          & \multicolumn{1}{c}{Innermost storage} \\ \hline

\\[-0.9em]
\multicolumn{4}{l}{\textbf{(b)}}                                                                                                                   \\ 
\multicolumn{1}{c|}{\multirow{2}{*}{\textbf{SAFs Choices}}}     & \multicolumn{3}{c}{\textbf{Operand Intersection}}                                       \\ \cline{2-4} 
\multicolumn{1}{c|}{}                                  & \multicolumn{1}{c|}{Off-chip} & \multicolumn{2}{c}{On-chip}                    \\ \hline 
\multicolumn{1}{c|}{\textbf{InnermostSkip}}                     & \multicolumn{1}{c|}{None}     & \multicolumn{2}{c}{ $SkipB\leftrightarrow A$}                       \\ \hline
\multicolumn{1}{c|}{\textbf{HierarchicalSkip}}                  & \multicolumn{1}{c|}{ $SkipB\leftrightarrow A$}     & \multicolumn{2}{c}{ $SkipB\leftrightarrow A$}                       \\ \hline
\end{tabular}

\caption{Choices for different design aspects: (a) dataflows (b) SAFs (representation formats and other minor SAFs are identical and thus are not shown for simplicity)}
\label{tab:case_study_archs}
\end{table}

%% file: 07_related_work.tex
\section{Related Work}
There is ample prior work in modeling frameworks for tensor accelerator designs.
These models can be classified into two classes: \emph{cycle-level models} and \emph{analytical models}. 

\textbf{Cycle-level models} evaluate the detailed cycle-level behaviors of potential designs. Many of them assume a specific target platform, such as ASIC~\cite{magnet} or FPGA~\cite{DNNbuilder, DNNFPGA1, DNNFPGA2} and perform register-transfer-level (RTL) analysis, which includes low-level hardware details (\eg, pipeline stages). There are also platform-independent models, \eg, STONNE\cite{STONNE}, which perform cycle-level architectural analysis without RTL implementations. While these cycle-level models are very accurate, they hinder the exploration among the vast number of dataflows due to their long simulation time~\cite{gamma, marvel, timeloop-ispass}. Furthermore, cycle-level models are often not well parameterized in terms of architecture topology, employed SAFs, dataflow, etc. These assumptions adversely limit the explorable designs.

\textbf{Analytical models}~\cite{timeloop-ispass, procrustes, Asilomar, accelergy-ispass, cosa, scale-sim, maestro, nnest-islped, DNN_predictor} perform higher level analytical evaluations without considering per-cycle processing details of the design. Since these models work on abstracted hardware models, they are usually well parameterized and modularized to support a wider range of architecture designs. However, to the authors' knowledge, they either do not recognize the sparse workloads or SAFs at all \cite{timeloop-ispass, accelergy-ispass, cosa, scale-sim, maestro, nnest-islped, DNN_predictor}, or only target design-specific SAFs \cite{procrustes, Asilomar}. For example, Procrustes\cite{procrustes} supports modeling of B format for one operand only. That is, no prior work aims to flexibly model general sparse tensor accelerators with various SAFs applied. Since at each architecture level, different SAFs can introduce different amounts of savings and overhead, the lack of trade-off analysis for SAFs prevents designers from using such analytical models for design space exploration. 

%% file: 08_conclusion.tex
\section{Conclusion}
Sparse tensor accelerators are important for efficiently processing many popular workloads. However, the lack of a unified description language and a modeling infrastructure to enable exploration of various designs impedes further advances in this domain. This paper proposes a systematic classification of sparsity-aware acceleration techniques into three high-level sparse acceleration features (SAFs): representation format, gating, and skipping. Exploiting this classification, we develop \tool, an analytical modeling framework for sparse tensor accelerators. We further observe that the analyses of dataflow, SAFs, and micro-architecture are orthogonal to each other. Based on the orthogonality, we design \tool's internal analysis as three decoupled steps to keep its modeling complexity tractable. To balance modeling accuracy and simulation speed, \tool~uses statistical characterizations of tensors. 

\tool~is over 2000$\times$ faster than cycle-level simulations, and models well-known sparse tensor accelerators with accurate relative trends and 0.1\% to 8\% average error. With case studies, we demonstrate that \tool\ can be used in accelerator design flows to help designers to compare and explore various designs, identify performance bottlenecks (\eg, memory bandwidth), and reveal broad design insights (\eg, co-design of sparsity, SAF and dataflow).

%% file: 10_artifact_appendix.tex
%
%
%
%
%

\newpage
\appendix
\section{Artifact Appendix}
\subsection{Abstract}

In this artifact, we provide the source code of \tool, its energy estimation backend based on Accelergy~\cite{accelergy-iccad}, and input specifications to key experimental results presented in the paper. To allow easy reproduction, we provide a docker environment with all necessary dependencies, automated scripts, and a Jupyter notebook that includes detailed instructions on running the evaluations. The artifact can be executed with any X86-64 machine with docker support and more than 10GB of disk space.

\subsection{Artifact check-list (meta-information)}


{\small
\begin{itemize}
  \item {\bf Algorithm:} Analytical modeling of sparse tensor accelerator performance (energy and cycles).
  \item {\bf Program: } C++, python.
  \item {\bf Run-time environment:} Dockerfile.
  \item {\bf Hardware:} Any X86-64 machine.
  \item {\bf Output:} Plots or tables generated from scripts.
  \item {\bf Experiments:} Analytical modeling of various sparse tensor accelerators running various workloads.
  \item {\bf How much disk space required (approximately)?:} 10GB 
  \item {\bf How much time is needed to prepare workflow (approximately)?:} Less than 30min if directly pulling docker image; less than 2 hours if building docker from the source.  
  \item {\bf How much time is needed to complete experiments (approximately)?:} Less than 1 hour to finish running all experiments in the provided default mode.
  \item {\bf Publicly available?:} Yes
  \item {\bf Code licenses (if publicly available)?:} MIT
  \item {\bf Archived (provide DOI)?: } Yes, DOI 10.5281/zenodo.7027215
\end{itemize}
}

\subsection{Description}

\subsubsection{How to access}
The artifact is hosted both on github (\url{https://github.com/Accelergy-Project/micro22-sparseloop-artifact}) and on an archival repository with DOI 10.5281/zenodo.7027215 (\url{https://doi.org/10.5281/zenodo.7027215
}).





\subsection{Installation}

Since we provide a docker, the installation process mainly involves obtaining the docker image that contains the dependencies, the compiled Sparseloop, and the energy estimation backend. Please follow the provided instructions (\url{https://github.com/Accelergy-Project/micro22-sparseloop-artifact/blob/main/README.md}) to obtain and start the docker. 

\subsection{Evaluation and expected results}

We provide a jupyter notebook in \emph{workspace/2022.micro. artifact/notebook/artifact\_evaluations.ipynb} to guide through the evaluations. Please navigate to the notebook in your docker Jupyter notebook file structure GUI. 

Each cell in the notebook provides the background, instructions, and commands to run each evaluation with provided scripts. The evaluations include the following key results from the paper:
\begin{itemize}
    \item Comparison of performance and energy for accelerators supporting different representation formats (Fig.~\ref{fig:mov.diff_arch}).
    \item Validations on various sparse tensor accelerators (Fig.~\ref{fig:eyeriss_v2_validation}, Table~\ref{tab:v1_compression}, Fig.~\ref{fig:dual-side-tensor-core}, and the STC design in Fig.~\ref{fig:stc-study}.)
    \item Example design flow using \tool~to perform apples-to-apples comparison, identify design limitations, and explore various solutions to the limitation (Fig.~\ref{fig:stc-study}).
\end{itemize}

The output of each evaluation will either produce a figure or the content of a table. The easiest way to check validity is to compare the generated figure/table with the ones in the paper. However, raw results can also be accessed in the \emph{workspace/evaluation\_setups} folder.  Please note that we had to use energy estimation data based on public technology node instead of our proprietary technology node, so the exact data might not match for certain evaluation(s). We explicitly point out such cases in the notebook.

\subsection{Experiment customization}
The input specifications in the \emph{workspace/evaluation\_setups} folder can be updated to specify different hardware setups (\eg, different buffer sizes). Moreover, we also provide options in the scripts to enable map space search using \tool\ (\eg, \emph{{-{}-} use\_mapper} option can be enabled).
\\


\subsection{Methodology}

Submission, reviewing and badging methodology:

\begin{itemize}
  \item \url{https://www.acm.org/publications/policies/artifact-review-badging}
  \item \url{http://cTuning.org/ae/submission-20201122.html}
  \item \url{http://cTuning.org/ae/reviewing-20201122.html}
\end{itemize}

%% file: main.bbl

\begin{thebibliography}{67}


\ifx \showCODEN    \undefined \def \showCODEN     #1{\unskip}     \fi
\ifx \showDOI      \undefined \def \showDOI       #1{#1}\fi
\ifx \showISBNx    \undefined \def \showISBNx     #1{\unskip}     \fi
\ifx \showISBNxiii \undefined \def \showISBNxiii  #1{\unskip}     \fi
\ifx \showISSN     \undefined \def \showISSN      #1{\unskip}     \fi
\ifx \showLCCN     \undefined \def \showLCCN      #1{\unskip}     \fi
\ifx \shownote     \undefined \def \shownote      #1{#1}          \fi
\ifx \showarticletitle \undefined \def \showarticletitle #1{#1}   \fi
\ifx \showURL      \undefined \def \showURL       {\relax}        \fi
\providecommand\bibfield[2]{#2}
\providecommand\bibinfo[2]{#2}
\providecommand\natexlab[1]{#1}
\providecommand\showeprint[2][]{arXiv:#2}

\bibitem[\protect\citeauthoryear{Abadi, Barham, Chen, Chen, Davis, Dean, Devin,
  Ghemawat, Irving, Isard, Kudlur, Levenberg, Monga, Moore, Murray, Steiner,
  Tucker, Vasudevan, Warden, Wicke, Yu, and Zheng}{Abadi et~al\mbox{.}}{2016}]%
        {tensorflow}
\bibfield{author}{\bibinfo{person}{Mart\'{\i}n Abadi}, \bibinfo{person}{Paul
  Barham}, \bibinfo{person}{Jianmin Chen}, \bibinfo{person}{Zhifeng Chen},
  \bibinfo{person}{Andy Davis}, \bibinfo{person}{Jeffrey Dean},
  \bibinfo{person}{Matthieu Devin}, \bibinfo{person}{Sanjay Ghemawat},
  \bibinfo{person}{Geoffrey Irving}, \bibinfo{person}{Michael Isard},
  \bibinfo{person}{Manjunath Kudlur}, \bibinfo{person}{Josh Levenberg},
  \bibinfo{person}{Rajat Monga}, \bibinfo{person}{Sherry Moore},
  \bibinfo{person}{Derek~G. Murray}, \bibinfo{person}{Benoit Steiner},
  \bibinfo{person}{Paul Tucker}, \bibinfo{person}{Vijay Vasudevan},
  \bibinfo{person}{Pete Warden}, \bibinfo{person}{Martin Wicke},
  \bibinfo{person}{Yuan Yu}, {and} \bibinfo{person}{Xiaoqiang Zheng}.}
  \bibinfo{year}{2016}\natexlab{}.
\newblock \showarticletitle{TensorFlow: A System for Large-Scale Machine
  Learning}. In \bibinfo{booktitle}{\emph{Proceedings of the 12th USENIX
  Conference on Operating Systems Design and Implementation (OSDI)}}.
  \bibinfo{pages}{265–283}.
\newblock
\showISBNx{9781931971331}


\bibitem[\protect\citeauthoryear{Albericio, Judd, Hetherington, Aamodt, Jerger,
  and Moshovos}{Albericio et~al\mbox{.}}{2016}]%
        {cnvlutin}
\bibfield{author}{\bibinfo{person}{Jorge Albericio}, \bibinfo{person}{Patrick
  Judd}, \bibinfo{person}{Tayler Hetherington}, \bibinfo{person}{Tor Aamodt},
  \bibinfo{person}{Natalie~Enright Jerger}, {and} \bibinfo{person}{Andreas
  Moshovos}.} \bibinfo{year}{2016}\natexlab{}.
\newblock \showarticletitle{Cnvlutin: Ineffectual-Neuron-Free Deep Neural
  Network Computing}. In \bibinfo{booktitle}{\emph{Proceedings of the 43rd
  Annual International Symposium on Computer Architecture (ISCA)}}.
  \bibinfo{pages}{1--13}.
\newblock
\urldef\tempurl%
\url{https://doi.org/10.1109/ISCA.2016.11}
\showDOI{\tempurl}


\bibitem[\protect\citeauthoryear{B{\"o}rner, Ernst, and G{\"u}ttel}{B{\"o}rner
  et~al\mbox{.}}{2015}]%
        {banded-matrix-application}
\bibfield{author}{\bibinfo{person}{Ralph-Uwe B{\"o}rner},
  \bibinfo{person}{Oliver~G Ernst}, {and} \bibinfo{person}{Stefan G{\"u}ttel}.}
  \bibinfo{year}{2015}\natexlab{}.
\newblock \showarticletitle{Three-dimensional transient electromagnetic
  modelling using rational {K}rylov methods}.
\newblock \bibinfo{journal}{\emph{Geophysical Journal International (Geophys.
  J. Int)}} \bibinfo{volume}{202}, \bibinfo{number}{3} (\bibinfo{year}{2015}),
  \bibinfo{pages}{2025--2043}.
\newblock


\bibitem[\protect\citeauthoryear{Bouaziz, Tagliasacchi, and Pauly}{Bouaziz
  et~al\mbox{.}}{2013}]%
        {graphics}
\bibfield{author}{\bibinfo{person}{Sofien Bouaziz}, \bibinfo{person}{Andrea
  Tagliasacchi}, {and} \bibinfo{person}{Mark Pauly}.}
  \bibinfo{year}{2013}\natexlab{}.
\newblock \showarticletitle{Sparse Iterative Closest Point}. In
  \bibinfo{booktitle}{\emph{Proceedings of the 11th Eurographics/ACMSIGGRAPH
  Symposium on Geometry Processing (SGP)}}. \bibinfo{pages}{113--123}.
\newblock
\urldef\tempurl%
\url{https://doi.org/10.1111/cgf.12178}
\showDOI{\tempurl}


\bibitem[\protect\citeauthoryear{Bulu\c{c}, Fineman, Frigo, Gilbert, and
  Leiserson}{Bulu\c{c} et~al\mbox{.}}{2009}]%
        {CSB}
\bibfield{author}{\bibinfo{person}{Aydin Bulu\c{c}}, \bibinfo{person}{Jeremy~T.
  Fineman}, \bibinfo{person}{Matteo Frigo}, \bibinfo{person}{John~R. Gilbert},
  {and} \bibinfo{person}{Charles~E. Leiserson}.}
  \bibinfo{year}{2009}\natexlab{}.
\newblock \showarticletitle{Parallel Sparse Matrix-Vector and
  Matrix-Transpose-Vector Multiplication Using Compressed Sparse Blocks}. In
  \bibinfo{booktitle}{\emph{Proceedings of the 21st Annual Symposium on
  Parallelism in Algorithms and Architectures (SPAA)}}.
  \bibinfo{pages}{233–244}.
\newblock
\showISBNx{9781605586069}
\urldef\tempurl%
\url{https://doi.org/10.1145/1583991.1584053}
\showDOI{\tempurl}


\bibitem[\protect\citeauthoryear{Chatarasi, Kwon, Parashar, Pellauer, Krishna,
  and Sarkar}{Chatarasi et~al\mbox{.}}{2021}]%
        {marvel}
\bibfield{author}{\bibinfo{person}{Prasanth Chatarasi},
  \bibinfo{person}{Hyoukjun Kwon}, \bibinfo{person}{Angshuman Parashar},
  \bibinfo{person}{Michael Pellauer}, \bibinfo{person}{Tushar Krishna}, {and}
  \bibinfo{person}{Vivek Sarkar}.} \bibinfo{year}{2021}\natexlab{}.
\newblock \showarticletitle{Marvel: A Data-Centric Approach for Mapping Deep
  Learning Operators on Spatial Accelerators}.
\newblock \bibinfo{journal}{\emph{ACM Transactions on Architecture and Code
  Optimization (TACO)}} \bibinfo{volume}{19}, \bibinfo{number}{1}, Article
  \bibinfo{articleno}{6} (\bibinfo{year}{2021}), \bibinfo{numpages}{26}~pages.
\newblock
\showISSN{1544-3566}
\urldef\tempurl%
\url{https://doi.org/10.1145/3485137}
\showDOI{\tempurl}


\bibitem[\protect\citeauthoryear{Chen, Emer, and Sze}{Chen
  et~al\mbox{.}}{2016}]%
        {eyeriss-isca}
\bibfield{author}{\bibinfo{person}{Yu-Hsin Chen}, \bibinfo{person}{Joel Emer},
  {and} \bibinfo{person}{Vivienne Sze}.} \bibinfo{year}{2016}\natexlab{}.
\newblock \showarticletitle{Eyeriss: A Spatial Architecture for
  Energy-Efficient Dataflow for Convolutional Neural Networks}. In
  \bibinfo{booktitle}{\emph{Proceedings of the 43rd Annual International
  Symposium on Computer Architecture (ISCA)}}. \bibinfo{pages}{367--379}.
\newblock
\urldef\tempurl%
\url{https://doi.org/10.1109/ISCA.2016.40}
\showDOI{\tempurl}


\bibitem[\protect\citeauthoryear{Chen, Krishna, Emer, and Sze}{Chen
  et~al\mbox{.}}{2017}]%
        {eyeriss-v1}
\bibfield{author}{\bibinfo{person}{Yu-Hsin Chen}, \bibinfo{person}{Tushar
  Krishna}, \bibinfo{person}{Joel~S. Emer}, {and} \bibinfo{person}{Vivienne
  Sze}.} \bibinfo{year}{2017}\natexlab{}.
\newblock \showarticletitle{Eyeriss: An Energy-Efficient Reconfigurable
  Accelerator for Deep Convolutional Neural Networks}.
\newblock \bibinfo{journal}{\emph{IEEE Journal of Solid-State Circuits (JSSC)}}
  \bibinfo{volume}{52}, \bibinfo{number}{1} (\bibinfo{year}{2017}),
  \bibinfo{pages}{127--138}.
\newblock
\urldef\tempurl%
\url{https://doi.org/10.1109/JSSC.2016.2616357}
\showDOI{\tempurl}


\bibitem[\protect\citeauthoryear{Chen, Yang, Emer, and Sze}{Chen
  et~al\mbox{.}}{2019}]%
        {eyeriss-v2}
\bibfield{author}{\bibinfo{person}{Yu-Hsin Chen}, \bibinfo{person}{Tien-Ju
  Yang}, \bibinfo{person}{Joel Emer}, {and} \bibinfo{person}{Vivienne Sze}.}
  \bibinfo{year}{2019}\natexlab{}.
\newblock \showarticletitle{Eyeriss v2: A Flexible Accelerator for Emerging
  Deep Neural Networks on Mobile Devices}.
\newblock \bibinfo{journal}{\emph{IEEE Journal on Emerging and Selected Topics
  in Circuits and Systems (JETCAS)}} \bibinfo{volume}{9}, \bibinfo{number}{2}
  (\bibinfo{year}{2019}), \bibinfo{pages}{292--308}.
\newblock
\urldef\tempurl%
\url{https://doi.org/10.1109/JETCAS.2019.2910232}
\showDOI{\tempurl}


\bibitem[\protect\citeauthoryear{Choquette, Lee, Krashinsky, Balan, and
  Khailany}{Choquette et~al\mbox{.}}{2021}]%
        {stc-isscc}
\bibfield{author}{\bibinfo{person}{Jack Choquette}, \bibinfo{person}{Edward
  Lee}, \bibinfo{person}{Ronny Krashinsky}, \bibinfo{person}{Vishnu Balan},
  {and} \bibinfo{person}{Brucek Khailany}.} \bibinfo{year}{2021}\natexlab{}.
\newblock \showarticletitle{The A100 Datacenter GPU and Ampere Architecture}.
  In \bibinfo{booktitle}{\emph{Proceedings of the IEEE International
  Solid-State Circuits Conference (ISSCC)}}, Vol.~\bibinfo{volume}{64}.
  \bibinfo{pages}{48--50}.
\newblock
\urldef\tempurl%
\url{https://doi.org/10.1109/ISSCC42613.2021.9365803}
\showDOI{\tempurl}


\bibitem[\protect\citeauthoryear{Chou, Kjolstad, and Amarasinghe}{Chou
  et~al\mbox{.}}{2018}]%
        {compression-formats}
\bibfield{author}{\bibinfo{person}{Stephen Chou}, \bibinfo{person}{Fredrik
  Kjolstad}, {and} \bibinfo{person}{Saman Amarasinghe}.}
  \bibinfo{year}{2018}\natexlab{}.
\newblock \showarticletitle{Format Abstraction for Sparse Tensor Algebra
  Compilers}.
\newblock \bibinfo{journal}{\emph{Proceedings of the ACM on Programming
  Languages (OOPSLA)}}  \bibinfo{volume}{2}, Article \bibinfo{articleno}{123}
  (\bibinfo{year}{2018}), \bibinfo{numpages}{30}~pages.
\newblock
\showISSN{2475-1421}


\bibitem[\protect\citeauthoryear{Deng, Sui, Liao, Qian, and Yuan}{Deng
  et~al\mbox{.}}{2021}]%
        {GoSPA}
\bibfield{author}{\bibinfo{person}{Chunhua Deng}, \bibinfo{person}{Yang Sui},
  \bibinfo{person}{Siyu Liao}, \bibinfo{person}{Xuehai Qian}, {and}
  \bibinfo{person}{Bo Yuan}.} \bibinfo{year}{2021}\natexlab{}.
\newblock \showarticletitle{GoSPA: An Energy-efficient High-performance
  Globally Optimized SParse Convolutional Neural Network Accelerator}. In
  \bibinfo{booktitle}{\emph{Proceedings of the 48th Annual International
  Symposium on Computer Architecture (ISCA)}}. \bibinfo{pages}{1110--1123}.
\newblock
\urldef\tempurl%
\url{https://doi.org/10.1109/ISCA52012.2021.00090}
\showDOI{\tempurl}


\bibitem[\protect\citeauthoryear{Devlin, Chang, Lee, and Toutanova}{Devlin
  et~al\mbox{.}}{2018}]%
        {bert}
\bibfield{author}{\bibinfo{person}{Jacob Devlin}, \bibinfo{person}{Ming{-}Wei
  Chang}, \bibinfo{person}{Kenton Lee}, {and} \bibinfo{person}{Kristina
  Toutanova}.} \bibinfo{year}{2018}\natexlab{}.
\newblock \showarticletitle{{BERT:} Pre-training of Deep Bidirectional
  Transformers for Language Understanding}.
\newblock \bibinfo{journal}{\emph{arXiv:1810.04805}} (\bibinfo{year}{2018}).
\newblock


\bibitem[\protect\citeauthoryear{{Einstein}}{{Einstein}}{1916}]%
        {einsum}
\bibfield{author}{\bibinfo{person}{A. {Einstein}}.}
  \bibinfo{year}{1916}\natexlab{}.
\newblock \showarticletitle{The Foundation of the General Theory of
  Relativity}.
\newblock \bibinfo{journal}{\emph{Annalen der Physik}} \bibinfo{volume}{354},
  \bibinfo{number}{7} (\bibinfo{year}{1916}), \bibinfo{pages}{769--822}.
\newblock
\urldef\tempurl%
\url{https://doi.org/10.1002/andp.19163540702}
\showDOI{\tempurl}


\bibitem[\protect\citeauthoryear{Gondimalla, Chesnut, Thottethodi, and
  Vijaykumar}{Gondimalla et~al\mbox{.}}{2019}]%
        {SparTen}
\bibfield{author}{\bibinfo{person}{Ashish Gondimalla}, \bibinfo{person}{Noah
  Chesnut}, \bibinfo{person}{Mithuna Thottethodi}, {and} \bibinfo{person}{T.~N.
  Vijaykumar}.} \bibinfo{year}{2019}\natexlab{}.
\newblock \showarticletitle{SparTen: A Sparse Tensor Accelerator for
  Convolutional Neural Networks}. In \bibinfo{booktitle}{\emph{Proceedings of
  the 52nd Annual IEEE/ACM International Symposium on Microarchitecture
  (MICRO)}}. \bibinfo{pages}{151--165}.
\newblock
\showISBNx{9781450369381}
\urldef\tempurl%
\url{https://doi.org/10.1145/3352460.3358291}
\showDOI{\tempurl}


\bibitem[\protect\citeauthoryear{Han, Mao, and Dally}{Han
  et~al\mbox{.}}{2016}]%
        {deep-compression}
\bibfield{author}{\bibinfo{person}{Song Han}, \bibinfo{person}{Huizi Mao},
  {and} \bibinfo{person}{William~J Dally}.} \bibinfo{year}{2016}\natexlab{}.
\newblock \showarticletitle{Deep Compression: Compressing Deep Neural Networks
  with Pruning, Trained Quantization and Huffman Coding}. In
  \bibinfo{booktitle}{\emph{Proceedings of the International Conference on
  Learning Representations (ICLR)}}. \bibinfo{pages}{1--14}.
\newblock


\bibitem[\protect\citeauthoryear{He, Zhang, Ren, and Sun}{He
  et~al\mbox{.}}{2015}]%
        {ResNet}
\bibfield{author}{\bibinfo{person}{Kaiming He}, \bibinfo{person}{Xiangyu
  Zhang}, \bibinfo{person}{Shaoqing Ren}, {and} \bibinfo{person}{Jian Sun}.}
  \bibinfo{year}{2015}\natexlab{}.
\newblock \showarticletitle{Deep Residual Learning for Image Recognition}.
\newblock \bibinfo{journal}{\emph{arXiv: 1512.03385}} (\bibinfo{year}{2015}).
\newblock


\bibitem[\protect\citeauthoryear{Hegde, Asghari-Moghaddam, Pellauer, Crago,
  Jaleel, Solomonik, Emer, and Fletcher}{Hegde et~al\mbox{.}}{2019}]%
        {extensor}
\bibfield{author}{\bibinfo{person}{Kartik Hegde}, \bibinfo{person}{Hadi
  Asghari-Moghaddam}, \bibinfo{person}{Michael Pellauer}, \bibinfo{person}{Neal
  Crago}, \bibinfo{person}{Aamer Jaleel}, \bibinfo{person}{Edgar Solomonik},
  \bibinfo{person}{Joel Emer}, {and} \bibinfo{person}{Christopher~W.
  Fletcher}.} \bibinfo{year}{2019}\natexlab{}.
\newblock \showarticletitle{ExTensor: An Accelerator for Sparse Tensor
  Algebra}. In \bibinfo{booktitle}{\emph{Proceedings of the 52nd Annual
  IEEE/ACM International Symposium on Microarchitecture (MICRO)}}.
  \bibinfo{pages}{319--333}.
\newblock
\showISBNx{9781450369381}


\bibitem[\protect\citeauthoryear{Henry, Fekete, and McGuffin}{Henry
  et~al\mbox{.}}{2007}]%
        {social-networks-graph}
\bibfield{author}{\bibinfo{person}{Nathalie Henry},
  \bibinfo{person}{Jean-Daniel Fekete}, {and} \bibinfo{person}{Michael~J.
  McGuffin}.} \bibinfo{year}{2007}\natexlab{}.
\newblock \showarticletitle{NodeTrix: a Hybrid Visualization of Social
  Networks}.
\newblock \bibinfo{journal}{\emph{IEEE Transactions on Visualization and
  Computer Graphics (IEEE Trans Vis Comput Graph)}} \bibinfo{volume}{13},
  \bibinfo{number}{6} (\bibinfo{year}{2007}), \bibinfo{pages}{1302--1309}.
\newblock
\urldef\tempurl%
\url{https://doi.org/10.1109/TVCG.2007.70582}
\showDOI{\tempurl}


\bibitem[\protect\citeauthoryear{Howard, Zhu, Chen, Kalenichenko, Wang, Weyand,
  Andreetto, and Adam}{Howard et~al\mbox{.}}{2017}]%
        {mobilenets}
\bibfield{author}{\bibinfo{person}{Andrew~G. Howard}, \bibinfo{person}{Menglong
  Zhu}, \bibinfo{person}{Bo Chen}, \bibinfo{person}{Dmitry Kalenichenko},
  \bibinfo{person}{Weijun Wang}, \bibinfo{person}{Tobias Weyand},
  \bibinfo{person}{Marco Andreetto}, {and} \bibinfo{person}{Hartwig Adam}.}
  \bibinfo{year}{2017}\natexlab{}.
\newblock \showarticletitle{MobileNets: Efficient Convolutional Neural Networks
  for Mobile Vision Applications}.
\newblock \bibinfo{journal}{\emph{arXiv: 1704.04861}} (\bibinfo{year}{2017}).
\newblock


\bibitem[\protect\citeauthoryear{Huang, Kang, Dinh, Norell, Kalaiah, Demmel,
  Wawrzynek, and Shao}{Huang et~al\mbox{.}}{2021}]%
        {cosa}
\bibfield{author}{\bibinfo{person}{Qijing Huang}, \bibinfo{person}{Minwoo
  Kang}, \bibinfo{person}{Grace Dinh}, \bibinfo{person}{Thomas Norell},
  \bibinfo{person}{Aravind Kalaiah}, \bibinfo{person}{James Demmel},
  \bibinfo{person}{John Wawrzynek}, {and} \bibinfo{person}{Yakun~Sophia Shao}.}
  \bibinfo{year}{2021}\natexlab{}.
\newblock \showarticletitle{CoSA: Scheduling by Constrained Optimization for
  Spatial Accelerators}. In \bibinfo{booktitle}{\emph{Proceedings of the 48th
  Annual ACM/IEEE International Symposium on Computer Architecture (ISCA)}}.
  \bibinfo{pages}{554--566}.
\newblock
\urldef\tempurl%
\url{https://doi.org/10.1109/ISCA52012.2021.00050}
\showDOI{\tempurl}


\bibitem[\protect\citeauthoryear{Jang, Lee, Kim, Park, Ardestani, Choi, Kim,
  Kim, Yu, Abdel-Aziz, Park, Lee, Lee, Kim, Jung, Nam, Lim, Lee, Song, Kwon,
  Hassoun, Lim, and Choi}{Jang et~al\mbox{.}}{2021}]%
        {samsung-npu}
\bibfield{author}{\bibinfo{person}{Jun-Woo Jang}, \bibinfo{person}{Sehwan Lee},
  \bibinfo{person}{Dongyoung Kim}, \bibinfo{person}{Hyunsun Park},
  \bibinfo{person}{Ali~Shafiee Ardestani}, \bibinfo{person}{Yeongjae Choi},
  \bibinfo{person}{Channoh Kim}, \bibinfo{person}{Yoojin Kim},
  \bibinfo{person}{Hyeongseok Yu}, \bibinfo{person}{Hamzah Abdel-Aziz},
  \bibinfo{person}{Jun-Seok Park}, \bibinfo{person}{Heonsoo Lee},
  \bibinfo{person}{Dongwoo Lee}, \bibinfo{person}{Myeong~Woo Kim},
  \bibinfo{person}{Hanwoong Jung}, \bibinfo{person}{Heewoo Nam},
  \bibinfo{person}{Dongguen Lim}, \bibinfo{person}{Seungwon Lee},
  \bibinfo{person}{Joon-Ho Song}, \bibinfo{person}{Suknam Kwon},
  \bibinfo{person}{Joseph Hassoun}, \bibinfo{person}{SukHwan Lim}, {and}
  \bibinfo{person}{Changkyu Choi}.} \bibinfo{year}{2021}\natexlab{}.
\newblock \showarticletitle{Sparsity-Aware and Re-configurable NPU Architecture
  for Samsung Flagship Mobile SoC}. In \bibinfo{booktitle}{\emph{2021 ACM/IEEE
  48th Annual International Symposium on Computer Architecture (ISCA)}}.
  \bibinfo{pages}{15--28}.
\newblock
\urldef\tempurl%
\url{https://doi.org/10.1109/ISCA52012.2021.00011}
\showDOI{\tempurl}


\bibitem[\protect\citeauthoryear{Kao and Krishna}{Kao and Krishna}{2020}]%
        {gamma}
\bibfield{author}{\bibinfo{person}{Sheng-Chun Kao} {and}
  \bibinfo{person}{Tushar Krishna}.} \bibinfo{year}{2020}\natexlab{}.
\newblock \showarticletitle{GAMMA: Automating the HW Mapping of DNN Models on
  Accelerators via Genetic Algorithm}. In \bibinfo{booktitle}{\emph{Proceedings
  of the IEEE/ACM International Conference On Computer Aided Design (ICCAD)}}.
  \bibinfo{pages}{1--9}.
\newblock


\bibitem[\protect\citeauthoryear{Ke, He, and Zhang}{Ke et~al\mbox{.}}{2018}]%
        {nnest-islped}
\bibfield{author}{\bibinfo{person}{Liu Ke}, \bibinfo{person}{Xin He}, {and}
  \bibinfo{person}{Xuan Zhang}.} \bibinfo{year}{2018}\natexlab{}.
\newblock \showarticletitle{NNest: Early-Stage Design Space Exploration Tool
  for Neural Network Inference Accelerators}. In
  \bibinfo{booktitle}{\emph{Proceedings of the International Symposium on Low
  Power Electronics and Design (ISLPED)}}. Article \bibinfo{articleno}{4},
  \bibinfo{numpages}{6}~pages.
\newblock
\showISBNx{9781450357043}
\urldef\tempurl%
\url{https://doi.org/10.1145/3218603.3218647}
\showDOI{\tempurl}


\bibitem[\protect\citeauthoryear{Khairy, Shen, Aamodt, and Rogers}{Khairy
  et~al\mbox{.}}{2020}]%
        {accelsim}
\bibfield{author}{\bibinfo{person}{Mahmoud Khairy}, \bibinfo{person}{Zhesheng
  Shen}, \bibinfo{person}{Tor~M. Aamodt}, {and} \bibinfo{person}{Timothy~G.
  Rogers}.} \bibinfo{year}{2020}\natexlab{}.
\newblock \showarticletitle{Accel-Sim: An Extensible Simulation Framework for
  Validated GPU Modeling}. In \bibinfo{booktitle}{\emph{Proceedings of the 47th
  ACM/IEEE Annual International Symposium on Computer Architecture (ISCA)}}.
  \bibinfo{pages}{473--486}.
\newblock
\urldef\tempurl%
\url{https://doi.org/10.1109/ISCA45697.2020.00047}
\showDOI{\tempurl}


\bibitem[\protect\citeauthoryear{Kjolstad, Chou, Lugato, Kamil, and
  Amarasinghe}{Kjolstad et~al\mbox{.}}{2017}]%
        {taco}
\bibfield{author}{\bibinfo{person}{Fredrik Kjolstad}, \bibinfo{person}{Stephen
  Chou}, \bibinfo{person}{David Lugato}, \bibinfo{person}{Shoaib Kamil}, {and}
  \bibinfo{person}{Saman Amarasinghe}.} \bibinfo{year}{2017}\natexlab{}.
\newblock \showarticletitle{Taco: A tool to generate tensor algebra kernels}.
  In \bibinfo{booktitle}{\emph{Proceedings of the 32nd IEEE/ACM International
  Conference on Automated Software Engineering (ASE)}}.
  \bibinfo{pages}{943--948}.
\newblock
\urldef\tempurl%
\url{https://doi.org/10.1109/ASE.2017.8115709}
\showDOI{\tempurl}


\bibitem[\protect\citeauthoryear{Kolodziej, Mahmoudi~Aznaveh, Bullock, David,
  Davis, Henderson, Hu, and Sandstrom}{Kolodziej et~al\mbox{.}}{2019}]%
        {suitesparse}
\bibfield{author}{\bibinfo{person}{Scott Kolodziej}, \bibinfo{person}{Mohsen
  Mahmoudi~Aznaveh}, \bibinfo{person}{Matthew Bullock},
  \bibinfo{person}{Jarrett David}, \bibinfo{person}{Timothy Davis},
  \bibinfo{person}{Matthew Henderson}, \bibinfo{person}{Yifan Hu}, {and}
  \bibinfo{person}{Read Sandstrom}.} \bibinfo{year}{2019}\natexlab{}.
\newblock \showarticletitle{The SuiteSparse Matrix Collection Website
  Interface}.
\newblock \bibinfo{journal}{\emph{Journal of Open Source Software (J. Open
  Source Softw.)}}  \bibinfo{volume}{4} (\bibinfo{year}{2019}),
  \bibinfo{pages}{1--4}.
\newblock
\urldef\tempurl%
\url{https://doi.org/10.21105/joss.01244}
\showDOI{\tempurl}


\bibitem[\protect\citeauthoryear{Krizhevsky, Sutskever, and Hinton}{Krizhevsky
  et~al\mbox{.}}{2012}]%
        {alexnet}
\bibfield{author}{\bibinfo{person}{Alex Krizhevsky}, \bibinfo{person}{Ilya
  Sutskever}, {and} \bibinfo{person}{Geoffrey~E Hinton}.}
  \bibinfo{year}{2012}\natexlab{}.
\newblock \showarticletitle{ImageNet Classification with Deep Convolutional
  Neural Networks}. In \bibinfo{booktitle}{\emph{Proceedings of Advances in
  Neural Information Processing Systems (NIPS)}}, Vol.~\bibinfo{volume}{25}.
\newblock


\bibitem[\protect\citeauthoryear{{Kwon}, {Chatarasi}, {Sarkar}, {Krishna},
  {Pellauer}, and {Parashar}}{{Kwon} et~al\mbox{.}}{2020}]%
        {maestro}
\bibfield{author}{\bibinfo{person}{H. {Kwon}}, \bibinfo{person}{P.
  {Chatarasi}}, \bibinfo{person}{V. {Sarkar}}, \bibinfo{person}{T. {Krishna}},
  \bibinfo{person}{M. {Pellauer}}, {and} \bibinfo{person}{A. {Parashar}}.}
  \bibinfo{year}{2020}\natexlab{}.
\newblock \showarticletitle{MAESTRO: A Data-Centric Approach to Understand
  Reuse, Performance, and Hardware Cost of DNN Mappings}.
\newblock \bibinfo{journal}{\emph{IEEE Micro}} \bibinfo{volume}{40},
  \bibinfo{number}{3} (\bibinfo{year}{2020}), \bibinfo{pages}{20--29}.
\newblock


\bibitem[\protect\citeauthoryear{Matr{\'i}nez, Abell{\'a}n, Acacio, and
  Krishna}{Matr{\'i}nez et~al\mbox{.}}{2021}]%
        {STONNE}
\bibfield{author}{\bibinfo{person}{Francisco~Mu{\~n}oz Matr{\'i}nez},
  \bibinfo{person}{Jos{\'e}~L. Abell{\'a}n}, \bibinfo{person}{Manuel~E.
  Acacio}, {and} \bibinfo{person}{Tushar Krishna}.}
  \bibinfo{year}{2021}\natexlab{}.
\newblock \showarticletitle{STONNE: Enabling Cycle-Level Microarchitectural
  Simulation for DNN Inference Accelerators}.
\newblock \bibinfo{journal}{\emph{IEEE Computer Architecture Letters (CAL)}}
  \bibinfo{number}{01} (\bibinfo{year}{2021}).
\newblock


\bibitem[\protect\citeauthoryear{Mei, Houshmand, Jain, Giraldo, and
  Verhelst}{Mei et~al\mbox{.}}{2020}]%
        {Zigzag}
\bibfield{author}{\bibinfo{person}{Linyan Mei}, \bibinfo{person}{Pouya
  Houshmand}, \bibinfo{person}{Vikram Jain}, \bibinfo{person}{Juan Sebastian~P.
  Giraldo}, {and} \bibinfo{person}{Marian Verhelst}.}
  \bibinfo{year}{2020}\natexlab{}.
\newblock \showarticletitle{ZigZag: {A} Memory-Centric Rapid {DNN} Accelerator
  Design Space Exploration Framework}.
\newblock \bibinfo{journal}{\emph{arxiv:2007.11360}} (\bibinfo{year}{2020}).
\newblock


\bibitem[\protect\citeauthoryear{{Motamedi}, {Gysel}, {Akella}, and
  {Ghiasi}}{{Motamedi} et~al\mbox{.}}{2016}]%
        {DNNFPGA1}
\bibfield{author}{\bibinfo{person}{M. {Motamedi}}, \bibinfo{person}{P.
  {Gysel}}, \bibinfo{person}{V. {Akella}}, {and} \bibinfo{person}{S.
  {Ghiasi}}.} \bibinfo{year}{2016}\natexlab{}.
\newblock \showarticletitle{Design space exploration of FPGA-based Deep
  Convolutional Neural Networks}. In \bibinfo{booktitle}{\emph{Proceedings of
  the 21st Asia and South Pacific Design Automation Conference (ASP-DAC)}}.
  \bibinfo{pages}{575--580}.
\newblock


\bibitem[\protect\citeauthoryear{{Nikolić}, {Mahmoud}, and
  {Moshovos}}{{Nikolić} et~al\mbox{.}}{2018}]%
        {ineffectual-work}
\bibfield{author}{\bibinfo{person}{M. {Nikolić}}, \bibinfo{person}{M.
  {Mahmoud}}, {and} \bibinfo{person}{A. {Moshovos}}.}
  \bibinfo{year}{2018}\natexlab{}.
\newblock \showarticletitle{Characterizing Sources of Ineffectual Computations
  in Deep Learning Networks}. In \bibinfo{booktitle}{\emph{Proceedings of the
  IEEE International Symposium on Workload Characterization (IISWC)}}.
  \bibinfo{pages}{86--87}.
\newblock
\urldef\tempurl%
\url{https://doi.org/10.1109/IISWC.2018.8573509}
\showDOI{\tempurl}


\bibitem[\protect\citeauthoryear{NVIDIA}{NVIDIA}{2020}]%
        {ampere-white-paper}
\bibfield{author}{\bibinfo{person}{NVIDIA}.} \bibinfo{year}{2020}\natexlab{}.
\newblock \bibinfo{booktitle}{\emph{NVIDIA A100 Tensor Core GPU Architecture}}.
\newblock \bibinfo{type}{{T}echnical {R}eport}. \bibinfo{institution}{NVIDIA}.
\newblock


\bibitem[\protect\citeauthoryear{{Pal}, {Beaumont}, {Park}, {Amarnath}, {Feng},
  {Chakrabarti}, {Kim}, {Blaauw}, {Mudge}, and {Dreslinski}}{{Pal}
  et~al\mbox{.}}{2018}]%
        {OuterSPACE}
\bibfield{author}{\bibinfo{person}{S. {Pal}}, \bibinfo{person}{J. {Beaumont}},
  \bibinfo{person}{D. {Park}}, \bibinfo{person}{A. {Amarnath}},
  \bibinfo{person}{S. {Feng}}, \bibinfo{person}{C. {Chakrabarti}},
  \bibinfo{person}{H. {Kim}}, \bibinfo{person}{D. {Blaauw}},
  \bibinfo{person}{T. {Mudge}}, {and} \bibinfo{person}{R. {Dreslinski}}.}
  \bibinfo{year}{2018}\natexlab{}.
\newblock \showarticletitle{OuterSPACE: An Outer Product Based Sparse Matrix
  Multiplication Accelerator}. In \bibinfo{booktitle}{\emph{Proceedings of the
  International Symposium on High Performance Computer Architecture (HPCA)}}.
  \bibinfo{pages}{724--736}.
\newblock
\urldef\tempurl%
\url{https://doi.org/10.1109/HPCA.2018.00067}
\showDOI{\tempurl}


\bibitem[\protect\citeauthoryear{Parashar, Raina, Shao, Chen, Ying, Mukkara,
  Venkatesan, Khailany, Keckler, and Emer}{Parashar et~al\mbox{.}}{2019}]%
        {timeloop-ispass}
\bibfield{author}{\bibinfo{person}{Angshuman Parashar},
  \bibinfo{person}{Priyanka Raina}, \bibinfo{person}{Yakun~Sophia Shao},
  \bibinfo{person}{Yu-Hsin Chen}, \bibinfo{person}{Victor~A. Ying},
  \bibinfo{person}{Anurag Mukkara}, \bibinfo{person}{Rangharajan Venkatesan},
  \bibinfo{person}{Brucek Khailany}, \bibinfo{person}{Stephen~W. Keckler},
  {and} \bibinfo{person}{Joel Emer}.} \bibinfo{year}{2019}\natexlab{}.
\newblock \showarticletitle{Timeloop: A Systematic Approach to DNN Accelerator
  Evaluation}. In \bibinfo{booktitle}{\emph{2019 IEEE International Symposium
  on Performance Analysis of Systems and Software (ISPASS)}}.
  \bibinfo{pages}{304--315}.
\newblock
\urldef\tempurl%
\url{https://doi.org/10.1109/ISPASS.2019.00042}
\showDOI{\tempurl}


\bibitem[\protect\citeauthoryear{Parashar, Rhu, Mukkara, Puglielli, Venkatesan,
  Khailany, Emer, Keckler, and Dally}{Parashar et~al\mbox{.}}{2017}]%
        {scnn}
\bibfield{author}{\bibinfo{person}{Angshuman Parashar}, \bibinfo{person}{Minsoo
  Rhu}, \bibinfo{person}{Anurag Mukkara}, \bibinfo{person}{Antonio Puglielli},
  \bibinfo{person}{Rangharajan Venkatesan}, \bibinfo{person}{Brucek Khailany},
  \bibinfo{person}{Joel Emer}, \bibinfo{person}{Stephen~W. Keckler}, {and}
  \bibinfo{person}{William~J. Dally}.} \bibinfo{year}{2017}\natexlab{}.
\newblock \showarticletitle{SCNN: An Accelerator for Compressed-Sparse
  Convolutional Neural Networks}. In \bibinfo{booktitle}{\emph{Proceedings of
  the 44th Annual International Symposium on Computer Architecture (ISCA)}}.
  \bibinfo{pages}{27--40}.
\newblock
\showISBNx{9781450348928}
\urldef\tempurl%
\url{https://doi.org/10.1145/3079856.3080254}
\showDOI{\tempurl}


\bibitem[\protect\citeauthoryear{Paszke, Gross, Massa, Lerer, Bradbury, Chanan,
  Killeen, Lin, Gimelshein, Antiga, Desmaison, Kopf, Yang, DeVito, Raison,
  Tejani, Chilamkurthy, Steiner, Fang, Bai, and Chintala}{Paszke
  et~al\mbox{.}}{2019}]%
        {pytorch}
\bibfield{author}{\bibinfo{person}{Adam Paszke}, \bibinfo{person}{Sam Gross},
  \bibinfo{person}{Francisco Massa}, \bibinfo{person}{Adam Lerer},
  \bibinfo{person}{James Bradbury}, \bibinfo{person}{Gregory Chanan},
  \bibinfo{person}{Trevor Killeen}, \bibinfo{person}{Zeming Lin},
  \bibinfo{person}{Natalia Gimelshein}, \bibinfo{person}{Luca Antiga},
  \bibinfo{person}{Alban Desmaison}, \bibinfo{person}{Andreas Kopf},
  \bibinfo{person}{Edward Yang}, \bibinfo{person}{Zachary DeVito},
  \bibinfo{person}{Martin Raison}, \bibinfo{person}{Alykhan Tejani},
  \bibinfo{person}{Sasank Chilamkurthy}, \bibinfo{person}{Benoit Steiner},
  \bibinfo{person}{Lu Fang}, \bibinfo{person}{Junjie Bai}, {and}
  \bibinfo{person}{Soumith Chintala}.} \bibinfo{year}{2019}\natexlab{}.
\newblock \showarticletitle{PyTorch: An Imperative Style, High-Performance Deep
  Learning Library}.
\newblock In \bibinfo{booktitle}{\emph{Proceedings of Advances in Neural
  Information Processing Systems (NIPS)}}. \bibinfo{pages}{8024--8035}.
\newblock


\bibitem[\protect\citeauthoryear{{Qin}, {Samajdar}, {Kwon}, {Nadella},
  {Srinivasan}, {Das}, {Kaul}, and {Krishna}}{{Qin} et~al\mbox{.}}{2020}]%
        {SIGMA}
\bibfield{author}{\bibinfo{person}{E. {Qin}}, \bibinfo{person}{A. {Samajdar}},
  \bibinfo{person}{H. {Kwon}}, \bibinfo{person}{V. {Nadella}},
  \bibinfo{person}{S. {Srinivasan}}, \bibinfo{person}{D. {Das}},
  \bibinfo{person}{B. {Kaul}}, {and} \bibinfo{person}{T. {Krishna}}.}
  \bibinfo{year}{2020}\natexlab{}.
\newblock \showarticletitle{SIGMA: A Sparse and Irregular GEMM Accelerator with
  Flexible Interconnects for DNN Training}. In
  \bibinfo{booktitle}{\emph{Proceedings of the International Symposium on High
  Performance Computer Architecture (HPCA)}}. \bibinfo{pages}{58--70}.
\newblock
\urldef\tempurl%
\url{https://doi.org/10.1109/HPCA47549.2020.00015}
\showDOI{\tempurl}


\bibitem[\protect\citeauthoryear{{Rahman}, {Oh}, {Lee}, and {Choi}}{{Rahman}
  et~al\mbox{.}}{2017}]%
        {DNNFPGA2}
\bibfield{author}{\bibinfo{person}{A. {Rahman}}, \bibinfo{person}{S. {Oh}},
  \bibinfo{person}{J. {Lee}}, {and} \bibinfo{person}{K. {Choi}}.}
  \bibinfo{year}{2017}\natexlab{}.
\newblock \showarticletitle{Design space exploration of FPGA accelerators for
  convolutional neural networks}. In \bibinfo{booktitle}{\emph{Proceedings of
  the Design, Automation Test in Europe Conference Exhibition (DATE)}}.
  \bibinfo{pages}{1147--1152}.
\newblock


\bibitem[\protect\citeauthoryear{Ravazzi, Tempo, and Dabbene}{Ravazzi
  et~al\mbox{.}}{2018}]%
        {sparse-graph}
\bibfield{author}{\bibinfo{person}{Chiara Ravazzi}, \bibinfo{person}{Roberto
  Tempo}, {and} \bibinfo{person}{Fabrizio Dabbene}.}
  \bibinfo{year}{2018}\natexlab{}.
\newblock \showarticletitle{Learning Influence Structure in Sparse Social
  Networks}.
\newblock \bibinfo{journal}{\emph{IEEE Transactions on Control of Network
  Systems}} \bibinfo{volume}{5}, \bibinfo{number}{4} (\bibinfo{year}{2018}),
  \bibinfo{pages}{1976--1986}.
\newblock
\urldef\tempurl%
\url{https://doi.org/10.1109/TCNS.2017.2781367}
\showDOI{\tempurl}


\bibitem[\protect\citeauthoryear{Saad}{Saad}{2011}]%
        {CSR}
\bibfield{author}{\bibinfo{person}{Yousef Saad}.}
  \bibinfo{year}{2011}\natexlab{}.
\newblock \bibinfo{booktitle}{\emph{Numerical Methods for Large Eigenvalue
  Problems}}.
\newblock \bibinfo{publisher}{Society for Industrial and Applied Mathematics}.
\newblock
\urldef\tempurl%
\url{https://doi.org/10.1137/1.9781611970739}
\showDOI{\tempurl}


\bibitem[\protect\citeauthoryear{Samajdar, Joseph, Zhu, Whatmough, Mattina, and
  Krishna}{Samajdar et~al\mbox{.}}{2020}]%
        {scale-sim}
\bibfield{author}{\bibinfo{person}{Ananda Samajdar},
  \bibinfo{person}{Jan~Moritz Joseph}, \bibinfo{person}{Yuhao Zhu},
  \bibinfo{person}{Paul Whatmough}, \bibinfo{person}{Matthew Mattina}, {and}
  \bibinfo{person}{Tushar Krishna}.} \bibinfo{year}{2020}\natexlab{}.
\newblock \showarticletitle{A Systematic Methodology for Characterizing
  Scalability of DNN Accelerators using SCALE-Sim}. In
  \bibinfo{booktitle}{\emph{Proceedings of the IEEE International Symposium on
  Performance Analysis of Systems and Software (ISPASS)}}.
  \bibinfo{pages}{58--68}.
\newblock
\urldef\tempurl%
\url{https://doi.org/10.1109/ISPASS48437.2020.00016}
\showDOI{\tempurl}


\bibitem[\protect\citeauthoryear{Sharma, Park, Mahajan, Amaro, Kim, Shao,
  Mishra, and Esmaeilzadeh}{Sharma et~al\mbox{.}}{2016}]%
        {DnnWeaver}
\bibfield{author}{\bibinfo{person}{Hardik Sharma}, \bibinfo{person}{Jongse
  Park}, \bibinfo{person}{Divya Mahajan}, \bibinfo{person}{Emmanuel Amaro},
  \bibinfo{person}{Joon~Kyung Kim}, \bibinfo{person}{Chenkai Shao},
  \bibinfo{person}{Asit Mishra}, {and} \bibinfo{person}{Hadi Esmaeilzadeh}.}
  \bibinfo{year}{2016}\natexlab{}.
\newblock \showarticletitle{From high-level deep neural models to FPGAs}. In
  \bibinfo{booktitle}{\emph{Proceedings of the 49th Annual IEEE/ACM
  International Symposium on Microarchitecture (MICRO)}}.
  \bibinfo{pages}{1--12}.
\newblock
\urldef\tempurl%
\url{https://doi.org/10.1109/MICRO.2016.7783720}
\showDOI{\tempurl}


\bibitem[\protect\citeauthoryear{Simonyan and Zisserman}{Simonyan and
  Zisserman}{2015}]%
        {VGG}
\bibfield{author}{\bibinfo{person}{Karen Simonyan} {and}
  \bibinfo{person}{Andrew Zisserman}.} \bibinfo{year}{2015}\natexlab{}.
\newblock \showarticletitle{Very Deep Convolutional Networks for Large-Scale
  Image Recognition}.
\newblock \bibinfo{journal}{\emph{arXiv:1409.1556}} (\bibinfo{year}{2015}).
\newblock


\bibitem[\protect\citeauthoryear{Smith and Karypis}{Smith and Karypis}{2015}]%
        {CSF}
\bibfield{author}{\bibinfo{person}{Shaden Smith} {and} \bibinfo{person}{George
  Karypis}.} \bibinfo{year}{2015}\natexlab{}.
\newblock \showarticletitle{Tensor-Matrix Products with a Compressed Sparse
  Tensor}. In \bibinfo{booktitle}{\emph{Proceedings of the 5th Workshop on
  Irregular Applications: Architectures and Algorithms (IA3)}}. Article
  \bibinfo{articleno}{5}, \bibinfo{numpages}{7}~pages.
\newblock
\showISBNx{9781450340014}
\urldef\tempurl%
\url{https://doi.org/10.1145/2833179.2833183}
\showDOI{\tempurl}


\bibitem[\protect\citeauthoryear{Srivastava, Jin, Liu, Albonesi, and
  Zhang}{Srivastava et~al\mbox{.}}{2020}]%
        {matraptor}
\bibfield{author}{\bibinfo{person}{Nitish Srivastava}, \bibinfo{person}{Hanchen
  Jin}, \bibinfo{person}{Jie Liu}, \bibinfo{person}{David Albonesi}, {and}
  \bibinfo{person}{Zhiru Zhang}.} \bibinfo{year}{2020}\natexlab{}.
\newblock \showarticletitle{MatRaptor: A Sparse-Sparse Matrix Multiplication
  Accelerator Based on Row-Wise Product}. In
  \bibinfo{booktitle}{\emph{Proceedings of the 53rd Annual IEEE/ACM
  International Symposium on Microarchitecture (MICRO)}}.
  \bibinfo{pages}{766--780}.
\newblock
\urldef\tempurl%
\url{https://doi.org/10.1109/MICRO50266.2020.00068}
\showDOI{\tempurl}


\bibitem[\protect\citeauthoryear{Srivastava, Rong, Barua, Feng, Cao, Zhang,
  Albonesi, Sarkar, Chen, Petersen, Lowney, Herr, Hughes, Mattson, and
  Dubey}{Srivastava et~al\mbox{.}}{2019}]%
        {T2STensor}
\bibfield{author}{\bibinfo{person}{Nitish Srivastava}, \bibinfo{person}{Hongbo
  Rong}, \bibinfo{person}{Prithayan Barua}, \bibinfo{person}{Guanyu Feng},
  \bibinfo{person}{Huanqi Cao}, \bibinfo{person}{Zhiru Zhang},
  \bibinfo{person}{David Albonesi}, \bibinfo{person}{Vivek Sarkar},
  \bibinfo{person}{Wenguang Chen}, \bibinfo{person}{Paul Petersen},
  \bibinfo{person}{Geoff Lowney}, \bibinfo{person}{Adam Herr},
  \bibinfo{person}{Christopher Hughes}, \bibinfo{person}{Timothy Mattson},
  {and} \bibinfo{person}{Pradeep Dubey}.} \bibinfo{year}{2019}\natexlab{}.
\newblock \showarticletitle{T2S-Tensor: Productively Generating
  High-Performance Spatial Hardware for Dense Tensor Computations}. In
  \bibinfo{booktitle}{\emph{Proceedings of the 27th Annual IEEE International
  Symposium on Field-Programmable Custom Computing Machines (FCCM)}}.
  \bibinfo{pages}{181--189}.
\newblock
\urldef\tempurl%
\url{https://doi.org/10.1109/FCCM.2019.00033}
\showDOI{\tempurl}


\bibitem[\protect\citeauthoryear{Sun, Li, Geng, Stuijk, and Corporaal}{Sun
  et~al\mbox{.}}{2022}]%
        {dissecting-stc}
\bibfield{author}{\bibinfo{person}{Wei Sun}, \bibinfo{person}{Ang Li},
  \bibinfo{person}{Tong Geng}, \bibinfo{person}{Sander Stuijk}, {and}
  \bibinfo{person}{Henk Corporaal}.} \bibinfo{year}{2022}\natexlab{}.
\newblock \showarticletitle{Dissecting Tensor Cores via Microbenchmarks:
  Latency, Throughput and Numerical Behaviors}.
\newblock \bibinfo{journal}{\emph{arXiv: 12206.02874}} (\bibinfo{year}{2022}).
\newblock


\bibitem[\protect\citeauthoryear{Sze, Chen, Yang, and Emer}{Sze
  et~al\mbox{.}}{2020}]%
        {dnn_syn_lec}
\bibfield{author}{\bibinfo{person}{Vivienne Sze}, \bibinfo{person}{Yu-Hsin
  Chen}, \bibinfo{person}{Tien-Ju Yang}, {and} \bibinfo{person}{Joel~S. Emer}.}
  \bibinfo{year}{2020}\natexlab{}.
\newblock \showarticletitle{Efficient Processing of Deep Neural Networks}.
\newblock \bibinfo{journal}{\emph{Synthesis Lectures on Computer Architecture}}
  \bibinfo{volume}{15}, \bibinfo{number}{2} (\bibinfo{year}{2020}),
  \bibinfo{pages}{1--341}.
\newblock
\urldef\tempurl%
\url{https://doi.org/10.2200/S01004ED1V01Y202004CAC050}
\showDOI{\tempurl}


\bibitem[\protect\citeauthoryear{{The SciPy community}}{{The SciPy
  community}}{[n. d.]}]%
        {COO}
\bibfield{author}{\bibinfo{person}{{The SciPy community}}.} \bibinfo{year}{[n.
  d.]}\natexlab{}.
\newblock \bibinfo{title}{SciPy documentation}.
\newblock
\newblock
\urldef\tempurl%
\url{https://docs.scipy.org/doc/scipy/reference/generated/scipy.sparse.coo_matrix.html}
\showURL{%
\tempurl}


\bibitem[\protect\citeauthoryear{{Venkatesan}, {Shao}, {Wang}, {Clemons},
  {Dai}, {Fojtik}, {Keller}, {Klinefelter}, {Pinckney}, {Raina}, {Zhang},
  {Zimmer}, {Dally}, {Emer}, {Keckler}, and {Khailany}}{{Venkatesan}
  et~al\mbox{.}}{2019}]%
        {magnet}
\bibfield{author}{\bibinfo{person}{R. {Venkatesan}}, \bibinfo{person}{Y.~S.
  {Shao}}, \bibinfo{person}{M. {Wang}}, \bibinfo{person}{J. {Clemons}},
  \bibinfo{person}{S. {Dai}}, \bibinfo{person}{M. {Fojtik}},
  \bibinfo{person}{B. {Keller}}, \bibinfo{person}{A. {Klinefelter}},
  \bibinfo{person}{N. {Pinckney}}, \bibinfo{person}{P. {Raina}},
  \bibinfo{person}{Y. {Zhang}}, \bibinfo{person}{B. {Zimmer}},
  \bibinfo{person}{W.~J. {Dally}}, \bibinfo{person}{J. {Emer}},
  \bibinfo{person}{S.~W. {Keckler}}, {and} \bibinfo{person}{B. {Khailany}}.}
  \bibinfo{year}{2019}\natexlab{}.
\newblock \showarticletitle{MAGNet: A Modular Accelerator Generator for Neural
  Networks}. In \bibinfo{booktitle}{\emph{Proceedings of the IEEE/ACM
  International Conference on Computer-Aided Design (ICCAD)}}.
  \bibinfo{pages}{1--8}.
\newblock


\bibitem[\protect\citeauthoryear{Wang, Zhang, Xie, Guo, Liu, and Leng}{Wang
  et~al\mbox{.}}{2021}]%
        {dual-side-tensor-core}
\bibfield{author}{\bibinfo{person}{Yang Wang}, \bibinfo{person}{Chen Zhang},
  \bibinfo{person}{Zhiqiang Xie}, \bibinfo{person}{Cong Guo},
  \bibinfo{person}{Yunxin Liu}, {and} \bibinfo{person}{Jingwen Leng}.}
  \bibinfo{year}{2021}\natexlab{}.
\newblock \showarticletitle{Dual-Side Sparse Tensor Core}. In
  \bibinfo{booktitle}{\emph{Proceedings of the 48th Annual International
  Symposium on Computer Architecture (ISCA)}}. \bibinfo{pages}{1083--1095}.
\newblock
\showISBNx{9781450390866}
\urldef\tempurl%
\url{https://doi.org/10.1109/ISCA52012.2021.00088}
\showDOI{\tempurl}


\bibitem[\protect\citeauthoryear{Wu, Emer, and Sze}{Wu et~al\mbox{.}}{2019}]%
        {accelergy-iccad}
\bibfield{author}{\bibinfo{person}{Yannan~Nellie Wu}, \bibinfo{person}{Joel~S.
  Emer}, {and} \bibinfo{person}{Vivienne Sze}.}
  \bibinfo{year}{2019}\natexlab{}.
\newblock \showarticletitle{Accelergy: An Architecture-Level Energy Estimation
  Methodology for Accelerator Designs}. In
  \bibinfo{booktitle}{\emph{Proceedings of the IEEE/ACM International
  Conference on Computer-Aided Design (ICCAD)}}. \bibinfo{pages}{1--8}.
\newblock
\urldef\tempurl%
\url{https://doi.org/10.1109/ICCAD45719.2019.8942149}
\showDOI{\tempurl}


\bibitem[\protect\citeauthoryear{Wu, Sze, and Emer}{Wu et~al\mbox{.}}{2020}]%
        {accelergy-ispass}
\bibfield{author}{\bibinfo{person}{Yannan~Nellie Wu}, \bibinfo{person}{Vivienne
  Sze}, {and} \bibinfo{person}{Joel~S. Emer}.} \bibinfo{year}{2020}\natexlab{}.
\newblock \showarticletitle{An Architecture-Level Energy and Area Estimator for
  Processing-In-Memory Accelerator Designs}. In
  \bibinfo{booktitle}{\emph{Proceedings of the IEEE International Symposium on
  Performance Analysis of Systems and Software (ISPASS)}}.
  \bibinfo{pages}{116--118}.
\newblock
\urldef\tempurl%
\url{https://doi.org/10.1109/ISPASS48437.2020.00024}
\showDOI{\tempurl}


\bibitem[\protect\citeauthoryear{Wu, Tsai, Parashar, Sze, and Emer}{Wu
  et~al\mbox{.}}{[n. d.]a}]%
        {sparseloop-website}
\bibfield{author}{\bibinfo{person}{Yannan~Nellie Wu}, \bibinfo{person}{Po-An
  Tsai}, \bibinfo{person}{Angshuman Parashar}, \bibinfo{person}{Vivienne Sze},
  {and} \bibinfo{person}{Joel~S. Emer}.} \bibinfo{year}{[n. d.]}\natexlab{a}.
\newblock \bibinfo{title}{Sparseloop website}.
\newblock
\newblock
\urldef\tempurl%
\url{http://sparseloop.mit.edu/}
\showURL{%
\tempurl}


\bibitem[\protect\citeauthoryear{Wu, Tsai, Parashar, Sze, and Emer}{Wu
  et~al\mbox{.}}{[n. d.]b}]%
        {timeloop-github}
\bibfield{author}{\bibinfo{person}{Yannan~Nellie Wu}, \bibinfo{person}{Po-An
  Tsai}, \bibinfo{person}{Angshuman Parashar}, \bibinfo{person}{Vivienne Sze},
  {and} \bibinfo{person}{Joel~S. Emer}.} \bibinfo{year}{[n. d.]}\natexlab{b}.
\newblock \bibinfo{title}{Timeloop code base}.
\newblock
\newblock
\urldef\tempurl%
\url{https://github.com/NVlabs/timeloop}
\showURL{%
\tempurl}


\bibitem[\protect\citeauthoryear{Xu, Zhang, Hao, Zhao, Zhang, Wang, Li, Guan,
  Chen, and Lin}{Xu et~al\mbox{.}}{2020}]%
        {AutoDNNchip}
\bibfield{author}{\bibinfo{person}{Pengfei Xu}, \bibinfo{person}{Xiaofan
  Zhang}, \bibinfo{person}{Cong Hao}, \bibinfo{person}{Yang Zhao},
  \bibinfo{person}{Yongan Zhang}, \bibinfo{person}{Yue Wang},
  \bibinfo{person}{Chaojian Li}, \bibinfo{person}{Zetong Guan},
  \bibinfo{person}{Deming Chen}, {and} \bibinfo{person}{Yingyan Lin}.}
  \bibinfo{year}{2020}\natexlab{}.
\newblock \showarticletitle{AutoDNNchip: An Automated DNN Chip Predictor and
  Builder for Both FPGAs and ASICs}. In \bibinfo{booktitle}{\emph{Proceedings
  of the ACM/SIGDA International Symposium on Field-Programmable Gate Arrays
  (FPGA)}}. \bibinfo{pages}{40--50}.
\newblock
\showISBNx{9781450370998}
\urldef\tempurl%
\url{https://doi.org/10.1145/3373087.3375306}
\showDOI{\tempurl}


\bibitem[\protect\citeauthoryear{Yang, Ghasemazar, Ren, Golub, Lemieux, and
  Lis}{Yang et~al\mbox{.}}{2020}]%
        {procrustes}
\bibfield{author}{\bibinfo{person}{Dingqing Yang}, \bibinfo{person}{Amin
  Ghasemazar}, \bibinfo{person}{Xiaowei Ren}, \bibinfo{person}{Maximilian
  Golub}, \bibinfo{person}{Guy Lemieux}, {and} \bibinfo{person}{Mieszko Lis}.}
  \bibinfo{year}{2020}\natexlab{}.
\newblock \showarticletitle{Procrustes: a Dataflow and Accelerator for Sparse
  Deep Neural Network Training}. In \bibinfo{booktitle}{\emph{Proceedings of
  the 53rd Annual IEEE/ACM International Symposium on Microarchitecture
  (MICRO)}}. \bibinfo{pages}{711--724}.
\newblock
\urldef\tempurl%
\url{https://doi.org/10.1109/MICRO50266.2020.00064}
\showDOI{\tempurl}


\bibitem[\protect\citeauthoryear{Yang, Chen, Emer, and Sze}{Yang
  et~al\mbox{.}}{2017}]%
        {Asilomar}
\bibfield{author}{\bibinfo{person}{Tien-Ju Yang}, \bibinfo{person}{Yu-Hsin
  Chen}, \bibinfo{person}{Joel Emer}, {and} \bibinfo{person}{Vivienne Sze}.}
  \bibinfo{year}{2017}\natexlab{}.
\newblock \showarticletitle{A method to estimate the energy consumption of deep
  neural networks}. In \bibinfo{booktitle}{\emph{Proceedings of the 51st
  Asilomar Conference on Signals, Systems, and Computers (Asilomar)}}.
  \bibinfo{pages}{1916--1920}.
\newblock
\urldef\tempurl%
\url{https://doi.org/10.1109/ACSSC.2017.8335698}
\showDOI{\tempurl}


\bibitem[\protect\citeauthoryear{Zhang, Attaluri, Emer, and Sanchez}{Zhang
  et~al\mbox{.}}{2021}]%
        {GAMMA-accelerator}
\bibfield{author}{\bibinfo{person}{Guowei Zhang}, \bibinfo{person}{Nithya
  Attaluri}, \bibinfo{person}{Joel~S. Emer}, {and} \bibinfo{person}{Daniel
  Sanchez}.} \bibinfo{year}{2021}\natexlab{}.
\newblock \showarticletitle{Gamma: Leveraging Gustavson’s Algorithm to
  Accelerate Sparse Matrix Multiplication}. In
  \bibinfo{booktitle}{\emph{Proceedings of the 26th ACM International
  Conference on Architectural Support for Programming Languages and Operating
  Systems (ASPLOS)}}. \bibinfo{pages}{687–701}.
\newblock
\showISBNx{9781450383172}
\urldef\tempurl%
\url{https://doi.org/10.1145/3445814.3446702}
\showDOI{\tempurl}


\bibitem[\protect\citeauthoryear{Zhang, Du, Zhang, Lan, Liu, Li, Guo, Chen, and
  Chen}{Zhang et~al\mbox{.}}{2016}]%
        {cambriconX}
\bibfield{author}{\bibinfo{person}{Shijin Zhang}, \bibinfo{person}{Zidong Du},
  \bibinfo{person}{Lei Zhang}, \bibinfo{person}{Huiying Lan},
  \bibinfo{person}{Shaoli Liu}, \bibinfo{person}{Ling Li}, \bibinfo{person}{Qi
  Guo}, \bibinfo{person}{Tianshi Chen}, {and} \bibinfo{person}{Yunji Chen}.}
  \bibinfo{year}{2016}\natexlab{}.
\newblock \showarticletitle{Cambricon-X: An accelerator for sparse neural
  networks}. In \bibinfo{booktitle}{\emph{2016 49th Annual IEEE/ACM
  International Symposium on Microarchitecture (MICRO)}}.
  \bibinfo{pages}{1--12}.
\newblock
\urldef\tempurl%
\url{https://doi.org/10.1109/MICRO.2016.7783723}
\showDOI{\tempurl}


\bibitem[\protect\citeauthoryear{{Zhang}, {Wang}, {Zhu}, {Lin}, {Xiong}, {Hwu},
  and {Chen}}{{Zhang} et~al\mbox{.}}{2018}]%
        {DNNbuilder}
\bibfield{author}{\bibinfo{person}{X. {Zhang}}, \bibinfo{person}{J. {Wang}},
  \bibinfo{person}{C. {Zhu}}, \bibinfo{person}{Y. {Lin}}, \bibinfo{person}{J.
  {Xiong}}, \bibinfo{person}{W. {Hwu}}, {and} \bibinfo{person}{D. {Chen}}.}
  \bibinfo{year}{2018}\natexlab{}.
\newblock \showarticletitle{DNNBuilder: an Automated Tool for Building
  High-Performance DNN Hardware Accelerators for FPGAs}. In
  \bibinfo{booktitle}{\emph{Proceedings of the IEEE/ACM International
  Conference on Computer-Aided Design (ICCAD)}}. \bibinfo{pages}{1--8}.
\newblock


\bibitem[\protect\citeauthoryear{Zhang, Wang, Han, and Dally}{Zhang
  et~al\mbox{.}}{2020}]%
        {sparch}
\bibfield{author}{\bibinfo{person}{Z. Zhang}, \bibinfo{person}{H. Wang},
  \bibinfo{person}{S. Han}, {and} \bibinfo{person}{W.~J. Dally}.}
  \bibinfo{year}{2020}\natexlab{}.
\newblock \showarticletitle{SpArch: Efficient Architecture for Sparse Matrix
  Multiplication}. In \bibinfo{booktitle}{\emph{Proceedings of the IEEE
  International Symposium on High Performance Computer Architecture (HPCA)}}.
  \bibinfo{pages}{261--274}.
\newblock
\urldef\tempurl%
\url{https://doi.org/10.1109/HPCA47549.2020.00030}
\showDOI{\tempurl}


\bibitem[\protect\citeauthoryear{Zhao, Panda, Sapatnekar, and Blaauw}{Zhao
  et~al\mbox{.}}{2002}]%
        {power-simulation}
\bibfield{author}{\bibinfo{person}{M. Zhao}, \bibinfo{person}{R.V. Panda},
  \bibinfo{person}{S.S. Sapatnekar}, {and} \bibinfo{person}{D. Blaauw}.}
  \bibinfo{year}{2002}\natexlab{}.
\newblock \showarticletitle{Hierarchical analysis of power distribution
  networks}.
\newblock \bibinfo{journal}{\emph{IEEE Transactions on Computer-Aided Design of
  Integrated Circuits and Systems (TCAD)}} \bibinfo{volume}{21},
  \bibinfo{number}{2} (\bibinfo{year}{2002}), \bibinfo{pages}{159--168}.
\newblock
\urldef\tempurl%
\url{https://doi.org/10.1109/43.980256}
\showDOI{\tempurl}


\bibitem[\protect\citeauthoryear{Zhao, Li, Wang, Xu, Zhang, and Lin}{Zhao
  et~al\mbox{.}}{2020}]%
        {DNN_predictor}
\bibfield{author}{\bibinfo{person}{Yang Zhao}, \bibinfo{person}{Chaojian Li},
  \bibinfo{person}{Yue Wang}, \bibinfo{person}{Pengfei Xu},
  \bibinfo{person}{Yongan Zhang}, {and} \bibinfo{person}{Yingyan Lin}.}
  \bibinfo{year}{2020}\natexlab{}.
\newblock \showarticletitle{DNN-Chip Predictor: An Analytical Performance
  Predictor for DNN Accelerators with Various Dataflows and Hardware
  Architectures}. In \bibinfo{booktitle}{\emph{Proceedings of the International
  Conference on Acoustics, Speech, and Signal Processing (ICASSP)}}.
  \bibinfo{pages}{1593--1597}.
\newblock
\urldef\tempurl%
\url{https://doi.org/10.1109/ICASSP40776.2020.9053977}
\showDOI{\tempurl}


\bibitem[\protect\citeauthoryear{Zhu, Zhang, Gu, and Xie}{Zhu
  et~al\mbox{.}}{2019}]%
        {vector-sparse-tensor-core}
\bibfield{author}{\bibinfo{person}{Maohua Zhu}, \bibinfo{person}{Tao Zhang},
  \bibinfo{person}{Zhenyu Gu}, {and} \bibinfo{person}{Yuan Xie}.}
  \bibinfo{year}{2019}\natexlab{}.
\newblock \showarticletitle{Sparse Tensor Core: Algorithm and Hardware
  Co-Design for Vector-Wise Sparse Neural Networks on Modern GPUs}. In
  \bibinfo{booktitle}{\emph{Proceedings of the 52nd Annual IEEE/ACM
  International Symposium on Microarchitecture (MICRO)}}.
  \bibinfo{pages}{359--371}.
\newblock
\showISBNx{9781450369381}
\urldef\tempurl%
\url{https://doi.org/10.1145/3352460.3358269}
\showDOI{\tempurl}


\end{thebibliography}
